# Analytical light scattering methods in molecular and structural biology: Experimental aspects and results

**Bernard LORBER**[*]

[a] Université de Strasbourg, CNRS, Architecture et Réactivité de l'ARN, UPR 9002 and Plateforme de Biophysique Esplanade ARBRE-MOBIEU[$], 15 rue René Descartes, 67084 Strasbourg, F-67000 Strasbourg, France

* Corresponding author: *E-mail address: b.lorber@ibmc-cnrs.unistra.fr* (B. Lorber)

*Postal address: Architecture et Réactivité de l'ARN, UPR 9002, Institut de Biologie Moléculaire du CNRS, 15 rue René Descartes, 67084 Strasbourg, France*

*Footnote:* $ ARBRE Association of Resources for Biophysical Research in Europe, MOBIEU MOlecular BIophysics in Europe.

## ABSTRACT

Non-invasive light scattering methods provide data on biological macromolecules (i.e. proteins, nucleic acids, as well as assemblies and larger entities composed of them) that are complementary with those of size exclusion chromatography, gel electrophoresis, analytical ultracentrifugation and mass spectrometry methods. Static light scattering measurements are useful to determine the mass of macromolecules and to monitor aggregation phenomena. Dynamic light scattering measurements are suitable for the quality control and to assess sample homogeneity, to determine particle size, examine the effect of physical and chemical treatments, probe the binding of ligands, and study interactions between macromolecules.

**Graphical abstract**

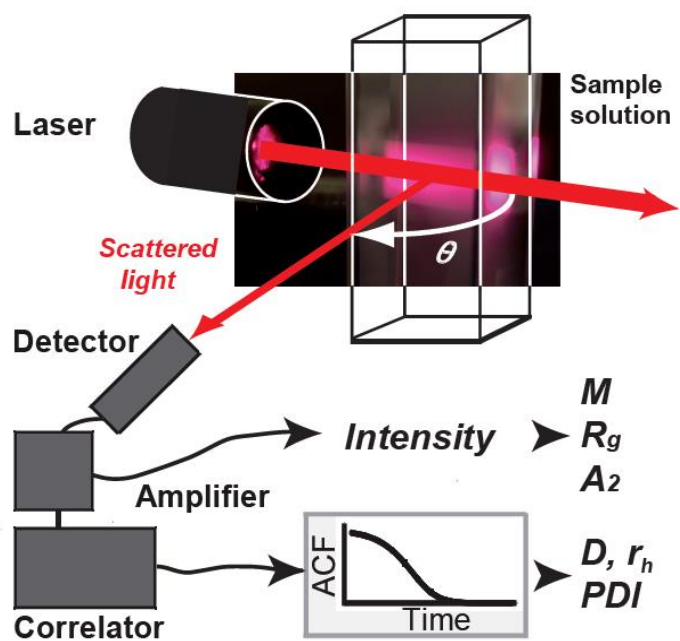

**Highlights**

- Light scattering methods are useful for the study of biological macromolecules
- Static light scattering is suitable to determine particle mass
- Dynamic light scattering is convenient to determine particle size

**Abbreviation used:** ACF, autocorrelation function; DLS, dynamic light scattering; HPLC, high performance liquid chromatography; kcps, $10^3$ counts/s; MALS, multi-angle light scattering; PSD, particle size distribution; SAXS, small-angle X-ray scattering; SEC, size exclusion chromatography; SLS, static light scattering.

## 1. Introduction

Analytical methods such as size exclusion chromatography, native gel electrophoresis, analytical ultracentrifugation or mass spectrometry, are essential to characterize, in terms of homogeneity, mass and size all pure biological macromolecules, including proteins, nucleic acids, assemblies made of proteins and/or nucleic acids, larger well-defined assemblies like ribosomes and viruses. Methods based on light scattering provide complementary data. Static and dynamic light scattering measurements are fast, non-invasive, and require only minute volumes of macromolecular solution. Static light scattering (SLS) exploits the proportionality relationship between the intensity of the light scattered and the mass and concentration of the macromolecule to derive the mass of the latter [1,2]. Dynamic light scattering (DLS, in which the word "dynamic" refers to objects moving freely in solution), records the fluctuations of the light scattered by the macromolecules as a consequence of Brownian motion and derives its size [3-5]. DLS is ideal to search for slowly diffusing particles (such as aggregates) and gives within seconds the size distribution in one microliter of solution [6].

The goal of this article is to convey, to students and researchers who are not familiar with light scattering methods, practical information about the application of the latter to determine reliable masses and sizes of macromolecules. Inevitably, the experimental aspects developed in detail may seem obvious to the specialist. A selection of original results obtained on a biophysics facility illustrate the wide array of applications the light scattering methods can have for the study of biological macromolecules and assemblies, as well as of systems composed of non-biological molecules.

## 2. Theoretical concepts

### 2.1. SLS

The Rayleigh-Debye-Zimm formalism expresses the mass of small isotropic particles as a function of the variation of the excess of scattered light intensity as:

$$Kc/R_\theta = (1/M\,P_\theta) + 2A_2c + \text{higher terms} \tag{1}$$

where $R_\theta$ is the Rayleigh ratio of scattered light to incident light, $M$ the mass of the particle, $c$ its concentration, and $A_2$ the second virial coefficient characterizing the interactions between particles. $K$ is an optical constant defined as:

$$K = (4\pi^2/\,\lambda^4 N_A)\,(n\,dn/dc)^2 \tag{2}$$

where $\lambda$ is the wavelength in vacuum, $n$ the refractive index of the solvent, $dn/dc$ the increment of refractive index of the macromolecule per concentration unit, and $N_A$ the Avogadro number. $P_\theta$ is related to particle size by:

$$1/P_\theta = 1 + (16\pi^2 n^2 R_g^2/3\lambda^2)\ sin^2(\theta/2) \tag{3}$$

where $R_g$ is the radius of gyration, i.e. the root mean square radius of the scattering elements, and $\theta$ the angle at which $I$ is measured. By convention, $\theta$ is measured clockwise starting from the light beam that exits the solution.

Intensity measurements at several angles and at various concentrations are necessary to determine simultaneously three properties of the dissolved particles using the Zimm plot representation [2]. (i) $R_g$ is derived from the slope of $Kc/(I_{measured}-I_{solvent})$ at various $\theta$ angles extrapolated at $c = 0$, (ii) $A_2$ is obtained from the slope of $Kc/(I_{measured}-I_{solvent})$ at various concentrations extrapolated at $\theta = 0$, and (iii) $1/M$ from intercept of the two previous graphs with the y axis. Size-exclusion chromatography separates, according to their hydrodynamic properties, the large particles from the interesting macromolecules producing a weaker signal before the detector measures $I$. Absorbance and refractive index measurements monitor macromolecular concentration $c$, required to calculate $M$ and $A_2$.

In single-angle SLS, as it is applied below, $P_\theta$ is assumed to be 1 for scatterers that are isotropic and small with respect to $\lambda$ (radius $\leq \lambda/10$). The Rayleigh-Debye-Zimm equation then reduces to

$$Kc/R_\theta = (1/M) + 2A_2c \tag{4}$$

A Debye plot representing $I$ as a function of $c$ is used to determine the mass of the particle by comparing $I$ (after subtraction of the intensity of the solvent) to that of toluene [7]. The graph of $Kc/R_\theta$ vs. $c$ extrapolates at $1/M$ for $c = 0$ with a slope equal to $A_2$.

The most performing commercially available MALS system measures $I$ at eighteen angles for the accurate determination of absolute mass and $A_2$. For most applications, an instrument doing measurements at three angles (e.g. at 49°, 90°, and 131°) is sufficient (**Fig. 1A**). In both cases, the macromolecule must be dilute to suppress multiple scattering. Small-angle X-ray scattering (SAXS) setups installed in laboratories or on synchrotron radiation sources are very similar to MALS setups, except that X-rays replace visible light and that the detector records the signal on a two-dimensional detector instead of a point. At low concentrations, the slope $Rg^2/3$ of the linear part of the Guinier plot $Ln [I(Q)] = f(Q)$, where $I(Q)$ is a function of the scattered intensity and Q the scattering variable, gives access to $Rg$ owing to an approximation [8].

### 2.2.    DLS

According to Einstein's treatment of diffusion in liquids, the mean square displacement $x^2$ of a molecule is related to its mutual diffusion coefficient $D$ and to time $t$ so that:

$$x^2 = 2Dt \qquad\qquad (5)$$

$D$ is inversely proportional to particle dimension and contains information about its size, shape and mass in a given medium and at a given temperature. Point scatterers with sizes below $\lambda/10$ produce Rayleigh scattering.

Dynamic light scattering (DLS) is a more recent designation for photon correlation spectroscopy (PCS) and for quasi-elastic light scattering (QELS), where quasi-elastic refers to the bouncing of photons when they collide with molecules or particles. DLS instruments are composed of a laser light source, a sample holder and a detector positioned at a fixed angle to collect photons in the plane of the laser beam (**Fig. 1B**). At short time intervals, a digital correlator compares the electronic current pulses converted to voltage pulses, fits the experimental data, builds an autocorrelation function that describes how the signal varies with time, and downloads the processed data onto a computer for further processing and display.

When a coherent light source illuminates a macromolecular solution, the intensity of the light scattered by the macromolecule fluctuates around a mean value. A speckle pattern is visible with the naked eye when very large slowly diffusing particles are present. The temporal fluctuations contain information about the dynamic process. The continuously changing distances between scatterers, result in constructive or destructive interferences by surrounding particles when the intensities add or subtract. The normalized second order autocorrelation function (ACF) of the scattered light intensity, i.e. the plot of the correlation coefficient vs the delay time $\tau$ generated by the correlator that describes particle motion then writes:

$$g^{(2)}(\tau) = <I(t)\ I(t+\tau)> / <I(t)^2> \qquad\qquad (6)$$

where $<I(t)>$ denotes the scattered total intensity at time $t$ and the brackets indicate the time average. If the time interval between intensity measurements is smaller than the delay time $\tau$ (also known as correlation or characteristic time), then a size distribution of the solute can be extracted from $\tau$. In the most general case, $g^{(2)}(\tau)$ follows the Siegert relation:

$$g^{(2)}(\tau) = B + \beta \times \{g^{(1)}(\tau)\}^2 \tag{7}$$

where $B$ is the baseline of the measurement (equal to 1 when all correlation is lost), $\beta$ a constant that depends on instrument geometry and optics, and $g^{(1)}(\tau)$ the normalized electric field time ACF that describes the dynamic process.

This ACF takes several forms depending on the composition of the sample as well as on the size and shape of the scatterers. The ACF of a population of identical particles decays exponentially and:

$$g^{(1)}(\tau) = exp\ (-\Gamma\tau) \tag{8}$$

where $\tau$ is the time relaxation of the decay and $\Gamma$ the decay rate:

$$\Gamma = Dq^2 \tag{9}$$

where $D$ is the mutual diffusion coefficient of the particles and $q$ the magnitude of the scattering wave vector equal to:

$$q = (4\pi n/\lambda)\ sin(\theta/2) \tag{10}$$

where $n$ is the refractive index of the solvent, $\lambda$ the wavelength in vacuum, and $\theta$ the scattering angle. $\Gamma$ is small at small $\theta$ angles. Hence, the relation between $\tau$ and $q$ is:

$$D = 1/2\ \tau q^2 \tag{11}$$

Particles with sizes smaller than $\sim\lambda/10$ (or $\sim60$ nm) are isotropic scattering centers that scatter light with the same intensity in all directions within the detection plane. On the opposite, particles with greater dimensions scatter more light forward, i.e. at small $\theta$ angles, and contribute more to the total intensity at $\theta \le 90°$.

Two mathematical methods extract particle size from DLS measurements. The method of cumulants [9] gives access to $\Gamma$ and hence to the distribution of decay rates of solutions containing a single population of particles. It either extrapolates the linear part onto the linear time axis (see **Supp.**

**Fig. 1A**) or fits estimates of the logarithm of the ACF to a polynomial function (see **Supp. Fig. 1B**). The latter approach assumes a Gaussian size distribution with a mean corresponding to the average size (calculated from first cumulant or moment) and a width corresponding to sample polydispersity (calculated from second cumulant) (see **Sup. Fig 1C**). The size distribution (PSD) of a population of identical particles is mono- (or uni) modal.

For a hard sphere, the Stokes-Einstein relation links the diffusion coefficient to the radius of a sphere. The hydrodynamic radius $r_h$ is the radius of the hard sphere that has the same diffusion coefficient. In the Stokes-Einstein relation

$$D = k_B T / 6\pi\eta r_h \tag{12}$$

$k_B$ is Boltzmann's constant (1.381 x $10^{-23}$ J/K) and $\eta$ the absolute (or dynamic) viscosity of the solvent. The expression at the denominator is the friction constant. $r_h$ is proportional to the time relaxation of the decay $\tau$ and to the inverse of $D$:

$$r_h = k_B T / 6\pi\eta D \tag{13}$$

For a sphere, $R_g/r_h \sim 0.77$. The $R_g$ of most hydrated macromolecules (e.g. proteins) is slightly smaller than the geometric (i.e. the dry) radius obtained by image analysis. Biologists may prefer to express particle sizes as hydrodynamic diameters, $d_h$, instead of radii, $r_h$.

If the diffusion coefficients of two populations of particles differ by a factor > 10, then PSD is bimodal and a single exponential function cannot fit the ACF. Smaller differences give a broad unimodal distribution. The mathematical inversion with the constrained regularization algorithm CONTIN [10] represents $g(\tau)$ by an integral over the distribution of normalized decay rates. Most DLS software apply to every measurement a Mie scattering function across the size range beyond 100 nm.

In extremely dilute (or ideal) solutions in which all molecules are soluble, the translational diffusion coefficient $D_0$ is equal to the mutual diffusion coefficient D. In non-ideal solutions, there exist inter-molecular interactions and $D$ decomposes as:

$$D = D_0 (1 + k_D c + ...) \tag{14}$$

where $k_D$ is the interaction parameter and $c$ the concentration. Very large negative values of $k_D$ are associated with attractive interactions.

## 3. Materials and methods

### 3.1. Light scattering instruments and software

Four instruments have produced the results displayed in the figures. They are: (i) a Protein Solutions DynaPro™ DP801 (20 mW He-Ne laser, $\lambda = 833$ nm, scattering angle $\theta = 90°$, 20 channel correlator), (ii) a Malvern Zetasizer™ NanoZS (4 mW He-Ne laser, $\lambda = 633$ nm, $\theta = 173°$ for backscattering measurements, correlator with 192 channels in eight groups of 24), (iii) a Wyatt Technology DynaPro Nanostar™ (100 mW He-Ne laser, $\lambda = 633$ nm, DLS $\theta = 90°$, 500 channel correlator, and equipped with a SLS detector at $\theta = 90°$) and (iv) a Wyatt MiniDawn TREOS ($\lambda = 658$ nm, $\theta = 49°$, 90° and 131°, and equipped for DLS at $\theta = 90°$). The latter is online with a HPLC-SEC column to separate the components of a sample prior to SLS measurements. The needed sample volumes are 20 µL in quartz cells with the DP-801 and the Zetasizer, and 1 µL with the Nanostar. The scattering volume of the flow cell of the TREOS instrument is less than 1 µL. The software operating with these instruments contain algorithms to do automated temperature ramps and data collection at time intervals.

The DLS instrument software represent the ACFs with different scales for the ordinate axis depending on the equation. The scale spans from e.g. 1.0 to 1.8 with the DP-801, from zero to one with the Zetaziser, and from one to 1.2 with the Nanostar. All autocorrelation curves displayed here have their ordinates normalized for the sake of homogeneity. Moreover, all PSDs shown hereafter represent the percentage of total $I$ (in counts) as a function of $d_h$ because those represented as percentage of total $I$ are not comparable since $I$ is set to 100% independently of the real number of counts.

### 3.2. Sample preparation

The users of the facility prepare their macromolecules. They calculate the refractive index and the absolute viscosity of their solvent using values found in a database, or measure them with the auxiliary instruments available at the facility (see procedures under supplementary material). In what follows, references indicate articles describing the preparation of the mentioned macromolecules and figure legends contain information about the solvents.

### 3.3. Light scattering measurements

The goal of preliminary measurements is to identify experimental conditions in which the response of the detector is proportional to macromolecular concentration (**Supp. Fig. 2**) using clean quartz cuvettes and solvent filtered through a membrane with a pore diameter of 0.1 µm and degassed with argon. Pure and dehydrated toluene serves as the universal reference in SLS [7]. Subtraction of the scattering intensity of the solvent filtered over a 20 nm pore membrane from that of the macromolecule is necessary. Filtration leads to the denaturation/aggregation of proteins because of constriction inside the pores of the filter and great pressure differences between inside and outside. Constant temperature and pH are critical for the reproducible preparation and analysis of macromolecules. A one hour-long micro-ultracentrifugation at 100,000 x $g$ at the temperature of the measurements removes particles with $d_h \leq 100$ nm. A low-speed centrifugation of the cuvettes filled with macromolecular solution (10 min at 500 x $g$ or 2200 rpm in a Sigma *1-6P* tabletop centrifuge), eliminates air bubbles and dust particles prior to measurements. At least ten DLS measurements in a row (with acquisition times of five to 15 s), are required to be sure that the signal is stable and the measurements are repeatable. Aggregates persisting after ultracentrifugation may form inside or outside the pipet tip during the transfer of the sample solution into the cuvette or at the contact with a solid surface. Solvent poured into the cuvette and removed quickly before introduction of the macromolecule can prevent aggregation. This step may be essential for delicate proteins.

SLS measurements at a single scattering angle imply to measure $I$ on a series of dilutions one after the other in the same cuvette and at a concentration that remains closest to the theoretical value. Gentle but insufficient mixing after dilution or a wrong extinction coefficient, result in errors on $c$ and hence on $M$. For a SEC-MALS analysis, ~100 micrograms protein (i.e. ~50 µL at 2 mg/mL) are loaded onto the chromatography column and eluted at 0.5 ml/min. Absorbance and/or refractive index are measured online for an accurate quantification of macromolecular concentration.

In SLS, the calculation of particle mass requires the solvent refractive index $n$ and the increment of refractive index with concentration $dn/dc$ of the dissolved particles (see above Eqs. 2 and 3). In DLS, the calculation of particle size takes into account the solvent refractive index $n$ and absolute solvent viscosity $\eta$ (see Eqs. 10 and 12). Any dissolved substance influences the refractive index $n$ and the viscosity $\eta$ of the solvent. $\eta$ varies more with temperature than $n$. **Supp. Fig 3** displays the properties of water and of glycerol and **Supp. Fig 4** highlights the effect of $n$ and $\eta$ corrections on a real PSD. Uncorrected data can easily lead to an error of a factor of two or three and hence to erroneous interpretations of $M$, oligomeric structure (e.g. from monomer to dimer or trimer) or shape (from spherical to very elongated). For the variety of biological macromolecules analyzed on the facility, the effect of the corrections for as many different solvents spans from "imperceptible" (when the properties of the ingredients of the solvent cancel each other out) to "substantial" (when the properties add). A few simple instruments help measure these values (see procedures under supplementary material).

### 3.4. Analysis and interpretation of results

Particle masses determined by SLS are realistic only if $I$ varies linearly with $c$ and if the macromolecular solutions do not contain larger scatterers that increase the scattering signal. As mentioned above, an experimental error on $c$ or on $I$ generates an error on $M$.

DLS measurements are performed to find out if a solution contains one or more populations of particles and to determine its mean $r_h$ or $d_h$ in the first case. Monomodal (or unimodal) macromolecular solutions are characterized by a single exponential ACF and a distribution fit that goes through all experimental points and reach rapidly the baseline (e.g. within $1.000 \pm 0.002$ as displayed by some software). In addition, $I$ varies by less than 10% from one measurement to the next. The sum of the squares of the deviations from the mean is low (e.g. between 0.1 and 5).

For many experimenters, the representation of the same size distribution according to intensity (I), volume (V), number (N), or mass (M) is confusing as to which one reflects the true composition of the sample. The intensity data do not imply any assumption about the applied Mie scattering function and are closest to reality. The transformation of an $I$ distribution into a V or a M distribution assumes that all particles have the same optical properties (which may be true) and the same shape (which may not be true). It minimizes the contribution of large particles in all cases and sometimes gives a false impression of mono- or unimodality. PSD distributions by N are close to the size distributions of particles under vacuum seen on transmission electron microscopy images. Besides, the $Z_{average}$ computed by the software is the $I$-weighted mean $r_h$ (or $d_h$) of the whole collection of particles composing a sample. It may be far from the size of the major fast diffusing component if slowly diffusing components are present.

Hereafter, all masses are expressed as relative masses, $M_r$. Sizes are mean hydrodynamic diameters $d_h$ (in nm) with a standard deviation or a polydispersity (in nm). $d_h$ is the diameter of a sphere that diffuses with the same speed as the particle under examination. Polydispersity is the width of the assumed Gaussian distribution derived from cumulant analysis. The index of polydispersity (PDI) is the weight average molecular weight divided by the number average molecular weight. The percentage of polydispersity (or relative polydispersity) is equal to the square root of the PDI multiplied by 100. Homogeneous macromolecular samples are composed of a single species of scatterers and their polydispersity is zero by definition.

The calculation of $M_r$ from $r_h$ or $d_h$ leads to an erroneous value if the particle's shape deviates from that of a sphere or d there are strong interactions between particles. Further, DLS software apply an empirical power law derived from a small set of proteins that are supposed to be globular [12]. For this reason, it is recommended to determine $r_h$ (or $d_h$) at various concentrations and to extrapolate to zero concentration. A MALS analysis is an alternative for it gives an absolute $M_r$.

Shear forces inside the SEC column may break down aggregates (as electrical fields may do during electrophoresis) and macromolecules may then appear more homogeneous. Therefore, it may be interesting to compare the result with that of measurements done in batch. Analytical methods based on other principles, such as ultracentrifugation and mass spectrometry may be helpful at this stage. The comparison between the dry mass derived from chemical composition and the mass of the hydrated particle informs about the particle's shape. DLS software calculate also a frictional coefficient ($f = 6\pi\eta r_h$ in Eq. 7) and/or a Perrin factor (i.e. the ratio of the frictional coefficient of a sphere having the same volume as the particle to that of a sphere of same $M_r$). Cryo-electron microscopy is a direct means to visualize hydrated particles.

### 3.5. Advantages and limitations

SLS and DLS analyses are not invasive as long as the wavelength of the laser is adapted to the color of the solute so that the light is not absorbed, and the experiment temperature is compatible with the stability of the macromolecule. In batch analyses, the macromolecules are in true solution conditions, i.e. they are not subjected to any forces that alter their size or mass distribution.

On the one hand, SLS experiments in a cuvette have the disadvantage that few large scatterers contribute strongly to the total intensity and lead to an overestimation of the particle mass. On the other hand, DLS cannot resolve particle populations whose diffusion coefficients differ by a factor below ten. It is not reasonable to extract PSDs from intensity ACF of mixtures of three or more populations of scatterers, even if the size differences between them exceed one order of magnitude. Titration experiments can provide valuable information about the association of small molecules and large macromolecules (see Result section).

### 3.6. From qualitative to quantitative data

Single-angle SLS measurements executed under best conditions yield reliable masses. For DLS, the 90° angle chosen by most instrument manufacturers is a good compromise for proteins and e.g. icosahedral viruses that are small with respect to λ and produce isotropic Rayleigh scattering. The comparison of above $M_r$ with that calculated from chemical composition and that of a sphere based on $r_h$ or $d_h$, suggests a type of oligomeric structure. In addition to measuring the $M_r$ of a macromolecule, SEC-MALS also estimates the proportion of every component of a mixture. The method is extremely useful to be sure that a complex forms between two or more macromolecules when DLS measurements do not detect a size variation.

The quality of DLS measurements is always better in term of particle size when working with homogeneous and monodisperse macromolecules. The quotient of the $M_r$ derived from $d_h$ on the $M_r$ derived from $I$ measurements is the shape factor. Values between two and three are common for elongated proteins. The comparison of the PSD by intensity with the PSD by mass (or volume) indicates the quantity of large scatterers present in a sample. When the amount by weight is less than 0.1%, it is generally negligible despite its potential impact on the ACF. A short centrifugation can bring it back to zero. Titration experiments can establish the stoichiometry of the association if the size difference between free and bound macromolecules is sufficient. Finally, DLS measurements offer the possibility to compare the thermal stability of wild type and mutated proteins, and the stabilizing effect of ligands on proteins.

## 4. Results and discussion

### 4.1. SLS

#### 4.1.2. Mass determination

The major application of single angle SLS is the determination of the mass of solute macromolecules or particles. On the one hand, the Debye method using $I$ measurements at several concentrations, gives good results with the bacterial nucleoprotein complex called transamidosome (see Fig. 8 of ref. [12]). On the other hand, SEC-MALS confirms that the plant protein PRORP-2 from *Arabidopsis thaliana* is a monomer with $M_r \sim 60,000$ but this is not clear at all in SEC alone because of the elongated shape of the molecule (**Fig. 2**). Analytical ultracentrifugation and SAXS data confirm this oligomeric structure [13].

#### 4.1.1. Aggregation phenomena

A small number of particles with great dimensions enhance strongly the scattering signal of a population of smaller scatterers since the intensity of the scattering is proportional to the sixth power of the particle diameter. This property is advantageous to track and monitor two types of aggregation phenomena. The first type discussed hereafter, includes (i) small molecules such as peptides that are insoluble under various experimental conditions, (ii) a gel forming polysaccharide and (iii) detergents micelles. The second type, discussed in section 4.2.1., includes aggregates that are either present or appear in solutions of biological macromolecules.

The measurement of the intensity of the light scattered by the synthetic antitumor peptides m2d and m3d [14], whose cell toxicity is not a linear function of concentration, is a straightforward means to estimate their solubility limit. In **Supp. Fig. 5A**, $I$ varies linearly only in a limited peptide concentration range in the cell culture medium and insoluble matter forms above a critical value. Similarly, the cellular response to phosphopeptide P140 issued from spliceosomal U1-70K snRNP protein and recognized by lupus CD4+ T cells, is proportional to $c$ only in the interval where the peptide is soluble [15]. In **Supp. Fig. 5B**, $I$ measurements also define the solubility limit of cathecol-rhodanine derivatives that inhibit specifically bacterial deoxyxylulose phosphate reductoisomerase [16].

The polysaccharide agarose dissolves completely in water above its melting temperature ($T_m$) and the resulting sol forms a reversible network below the gelling temperature $T_g$. $T_m$ and $T_g$ depend upon the length of the polymer chain, the nature of the chemical groups grafted on it and the solvent composition. **Supp. Fig 5C** shows how $I$ increases when a 0.4% (m/v) aqueous solution forms a gel at $T < T_g \sim 30°C$.

$I$ measurements are also a means to estimate the critical micellization concentration (cmc) at which micelles form in detergent solutions. Micelles are composed of a number of detergent monomers called the aggregation number. **Supp. Fig. 5D** is the graph of $I = f(c)$ for the non-ionic detergent octyl glucoside in water at 20°C. As with other methods, the cmc is a concentration interval, here from 25 to 30 mM in agreement with published data [17]. This method is as fast as manual surface tension measurements.

### 4.1.3. Estimation of extinction coefficients

The proportionality between the scattered intensity $I$, particle $M_r$ and $c$ is practical to estimate the extinction coefficient of a virus of known size. Grapevine Fan Leaf Virus has an icosahedral and quasi-spherical shape. Its diameter is very close to that Brome Mosaic Virus whose extinction coefficient is $E_{260\,nm} = 5.1$ mg/mL/cm. The intensity of a virus suspension having the same absorbance at 260 nm leads to $E_{260\,nm} \sim 9$ mg/mL/cm for GFLV. This value is close to that calculated from amino acid and nucleotide composition [18].

### 4.2. DLS

Contemporary DLS instruments measure particle sizes over three orders of magnitude, from one nanometer to 1000 nm, or covering roughly the dimension range encompassing that of small proteins to that of viruses and bacteria (**Supp. Fig. 6**). The array of applications is broad, from the detection of aggregates to the monitoring of the self-association of diverse biological macromolecules and the determination of particle sizes for the purpose of structural studies. Most of the time, the limits are set more by experimenter's imagination than by technical constraints.

### 4.2.1. Aggregation

More than 90% of the pure proteins analyzed for the first time by DLS at the facility are heterogeneous despite optimized purification protocols. At a first sight, this may seem contradictory with the fact that these proteins are pure according to electrophoresis, size-exclusion chromatography, and mass spectrometry criteria. In reality, these proteins do not contain any foreign macromolecules but frequently clusters composed of a few to thousands randomly associated macromolecules called aggregates. For this reason, their quality is not of satisfactory for accurate biophysical studies. Ref. [19] summarizes the various causes and effects of heterogeneity.

Supp. Fig. 7 illustrates how limited proteolysis reduces the heterogeneity of human mitochondrial tyrosyl-tRNA synthetase [20]. This result has urged the production of a genetically engineered protein deprived of its floppy C-terminal S4-like domain (subunit $M_r$ of 40,000, $d_h$~7,6nm). The homogeneity and compactness of this novel molecule favor the growth of well-ordered crystals whose diffraction quality has yielded a high-resolution 3D structure.

The detection of large scatterers during the aggregation of proteins that follows the dissociation of their subunits is practical to study their thermal stability. The transition temperature (Tm) of aspartyl-tRNA synthetases from E. coli (ecDRS) and from human mitochondria (hmDRS) increases in the presence of a synthetic analog of the catalytic intermediate (aspartyl-sulfamoyl adenosine, AspSA). In Fig. 3, the Tm of both proteins differ by 10°C, but only by 7°C when the ligand is present. Differential scanning fluorimetry analyzes confirm these results [21].

### 4.2.2. Size of quasi-spherical viruses

There are no interactions between capsids in dilute suspensions of pure viruses. In Fig. 4, pure Arabis Mosaic Virus (ArMV), Brome Mosaic Virus (BMV), Grapevine Fan Leaf Virus (GFLV), Turnip Yellow Mosaic Virus (TYMV) and Tomato Bushy Stunt Virus (TBSV) exhibit monomodal PSDs. The $d_h$ of their capsids ranges from 32 nm for BMV and ArMV to 33 nm for GFLV, and from 34 nm for TYMV to 37 nm for TBSV, respectively. The dimensions are slightly smaller under vacuum in the transmission electron microscope [23]. All these DLS analyzes take about ten times less solution volume and are ten times faster than analytical ultracentrifugation.

### 4.2.3. Oligomerization of a membrane protein

The ACFs and PDSs in Fig 5A and B show the effects of two non-ionic detergents on the solubility of the voltage-dependent anion-selective channel VDAC-34 involved in the translocation of transfer RNAs through the mitochondrial outer membrane [24]. In Fig. 5C, real intensities replace the percentage of total intensity on the Y-axis of the graph after subtraction of the scattering signal of the detergent micelles from that of the protein surrounded by micelles. In the presence of lauryl dimethyl amine oxide, the pure protein behaves as a homogeneous population with a mean $d_h$ ~5 nm (Fig. 5A-C) corresponding to the monomer (Fig. 5D). In the presence of octyl glucoside, VDAC-34 forms objects with $d_h$ ~15 nm that have the size of the hexamers visualized by atomic force microscopy [25]. Beside this, the PSDs of solutions of detergents that are used with membrane proteins in Supp. Fig. 8, show that the size of the micelles is well defined. This is astonishing because of the dynamic nature of the micelles, which are either prolate or oblate objects characterized by a limited lifetime during which the individual detergent molecules continuously exchange [26].

### 4.2.4. Asymmetrical macromolecules

Spherical biological macromolecules are exceptions. The shapes can be anything else, for instance ellipsoids, discs or donuts. The sliding clamps involved in chromosomes replication are homodimers with ring-like shapes with $d_h$ from 9 to 11 nm in DLS [27]. Most proteins, nucleic acids, and nucleoprotein complexes a core with loops or domains pointing towards the solvent or of multiple domains that are sometimes arranged without symmetry. They have concavities at their surface in which ligands bind. DLS detects a difference between the free molecules and the complex if their sizes differ sufficiently but even in this case, this may become difficult if the associating partners have complementary shapes. Native methionyl-tRNA synthetase from *E. coli* is a symmetrical homodimer with a subunit $M_r$ ~80,000. It catalyzes the activation of methionine in the presence of ATP and $Mg^{2+}$ ions and loads methionyl adenylate onto the 3' terminal adenine of the cognate transfer RNA. The cleavage of the C-terminal domain of the protein yields a fully active monomer with $M_r$ ~64,000 and $d_h = 6.5 \pm 1$ nm [28]. In **Fig. 6**, the formation of a complex with $d_h = 8 \pm 1$ nm is seen upon saturation with tRNA despite the shape complementarity.

In the case of a camel single domain antibody (sdAb) with $M_r$~15,000 that binds to Grapevine Fan Leaf Virus ($d_h$ ~32 nm, $M_r$ ~ 5 $10^6$), DLS detects an increase of the diameter of the capsid by 3 nm. This is enough to assess the formation of a complex and to establish that 60 molecules bind per capsid, in either titration or single addition experiments (**Fig. 7**). The 3D structure of the complex solved by cryo-electron microscopy indeed reveals 60 antibody molecules bound per capsid [29]. The increase of $d_h$ by respectively of 11 nm and 18 nm with sdABs substituted by Green Fluorescent Protein ($M_r$ ~27,000) or dimeric alkaline phosphate from *E. coli* ($M_r$ 94,000) is a further argument in favor of the association. The negative control of these investigations is a natural mutant of the virus not recognized by the sdAb.

In incremental titration experiments, three or more ligands bind to a macromolecule one after the other. The asparaginylation transamidosome of *Thermus thermophilus* is composed of a non-discriminating aspartyl-tRNA synthetase, an amidotransferase and tRNA$^{Asn}$. **Supp. Fig 9A** shows that none of the macromolecules participating in the complex is spherical. The transfer RNA has the shape of a boomerang, one protein resembles a parallelepiped and the other is elongated. In which order do the components associate? DLS shows that the two proteins have no affinity one for the other but that a ternary complex forms as soon as tRNA$^{Asn}$ is added in the presence of aspartic acid, $Mg^{2+}$ ions, ATP, and glutamine as a donor of $NH_2$ (see **Supp. Fig 9B**)[30]. This complex has a $D$~3.0 $10^{-7}$ cm$^2$/s and $d_h$ ~13.6 nm, corresponding to an equivalent sphere with $M_r$ ~300,000. According to the Debye graph obtained after single angle SLS measurements, $M_r \sim 400,000$ as compared with $\sim 300,000$ in analytical ultracentrifugation and $\sim 380,000$ in size exclusion chromatography. SAXS analyses indicate that the maximal $d_h \sim 18.5$ nm, $R_g \sim 5.5$ nm and $M_r \sim 325,000 \pm 50,000$. The crystallographic 3D structure of the complex reveals that it contains one additional dimeric aspartyl-tRNA synthetase molecule carrying two tRNA molecules and has a total $M_r \sim 550,000$ (**Supp. Fig 9C**)[31]. This implies that the process of crystallization traps a transient state and that the soluble complexes is very dynamic. Other DLS analyses show that in *Helicobacter pylori*, the tRNA associates first with the transamidase, and the aminoacylation and transamidation reactions occur once aspartyl-tRNA synthetase binds to this complex [32].

It may be challenging to demonstrate with DLS alone, that macromolecules with complementary shapes do associate. At variance with the case of methionyl-tRNA synthetase displayed in **Fig. 6**, the attempt to demonstrate that homodimeric human mitochondrial aspartyl-tRNA synthetase ($M_r \sim 2$ x 70,000) binds tRNA$^{Asp}$ from *E. coli* ($M_r \sim 25,000$) fails. In **Supp. Fig. 10**, the complex has not a much greater $d_h$ than the free enzyme although enzymology assays prove that one tRNA binds per monomer with a good affinity [21]. A MALS analysis would be more successful because of the $M_r$ difference of $\sim 40,000$.

### 4.2.5. Inter-molecular interactions

Light scattering is appropriate to investigate interactions between proteins [33]. A Debye graph with a positive slope indicates that there are interactions. Homodimeric human mitochondrial aspartyl-tRNA synthetase shares 43% sequence homology with its homolog from *E. coli* [21]. The $M_r$ of the polypeptide chains are $\sim 70,000$ and $\sim 66,000$, respectively. The $d_h$ of mitochondrial enzyme is significantly greater than that of the bacterial one at a concentration of 10 mg/mL (**Supp. Fig. 11A**). Does this mean that the enzymes differ really in size? In **Supp Fig. 11B**, $I$ measurements after sequential dilution confirm that the bacterial protein is the most soluble in this solvent. At $c = 0$, however, the graphs extrapolate to the same diffusion coefficient ($D_0 \sim 3.5 \ 10^{-7}$ and $\sim 3.4 \ 10^{-7}$ cm$^2$/s, respectively) meaning that both proteins have actually comparable sizes. Their $d_h = 9.3$ nm and 10 nm, are those of spherical proteins with respectively $M_r \sim 125,000$ and $M_r \sim 145,000$. These values are close to the 132,000 and 140,000 calculated from amino acid composition. The 3D crystallographic structures of both protein confirm that they have the same dimensions (**Supp. Fig. 11C**) [34]. These data illustrate that the interactions in a 10 mg/mL solution lead to the false impression that the mitochondrial enzyme has a $d_h$ of 11.7 nm, equivalent to that of a spherical protein with $M_r$ 210,000, that is in other words 50% greater than in reality.

## 5. Discussion and conclusion

Homogeneous macromolecules are of paramount importance to obtain reliable quantitative data, such as molecular masses and sizes. This holds also true for binding stoichiometries, affinity constants between molecules, or thermodynamics interaction parameters obtained using a variety of biochemical and biophysical approaches [33]. DLS is an attractive non-invasive analytical tool for the quality control, i.e. to verify the homogeneity of macromolecular preparations, before undertaking time-consuming and expensive biophysical studies [19]. Together with gel electrophoresis, size exclusion chromatography, ultracentrifugation and mass spectrometry, it is suitable for the fast analysis of proteins, nucleic acids, ribosomes and viruses in the molecular and structural biology laboratory. It reveals that many pure macromolecules are in reality heterogeneous [35]. The introduction of a SEC step prior to light or X-ray scattering measurements eliminates the effect of large aggregates, and provides more meaningful $R_g$ and $A_2$. As noted above, DLS measurements are a simple way to determine the size of biological and non-biological molecules and particles. They help identify solvents in which macromolecules are more stable, and e.g. in which carbon nanotubes are soluble [36]. DLS is rapid to ascertain the size of liposomes, vesicles (e.g. [37]) and exosomes (e.g. [38]). It is ideal to compare the solubility of mutated proteins, investigate transitions from native to denatured states, in parallel with circular dichroism measurements [36]. The only thing to bear in mind is that the DLS raw data require corrections to take into account macromolecule and solvent properties. Amongst the innovations in light scattering analysis are plate readers performing measurements on large numbers of samples, e.g. for the study of label-free protein-protein interactions in solution [39]. Zeta potentials derived from electrophoretic DLS measurements [40] as well as the combination with Raman spectroscopy allow a better characterization of protein aggregates in pharmaceutical formulations [41]. Ref. [42] lists several other recent applications.

## Declaration of interest

The author has no conflict of interest to report.

## Funding information

This work has benefited from a continued funding from Université de Strasbourg and CNRS.

## Author contribution

B.L. has performed all measurements, except SEC-MALS measurement done by the INSTRUCT at the IGBMC, Illkirch. He has analyzed and interpreted all results, written the article and prepared all figures.

## Acknowlegments

In acknowledge the grants received from Université de Strasbourg and CNRS for the purchase of light scattering instruments and the people who supported the latter, namely R. Giegé (former head of the Protein Crystallogenesis team), E. Westhof (former head of UPR9002), S. Muller (head of UPR3572, Immunopathologie et Chimie Thérapeutique, IBMC, Strasbourg), J.M. Reichart (formerly at UPR9022, Réponse immunitaire et développement chez les Insectes), Ph. Giegé (IBMP-CNRS, Strasbourg) and B. Bechinger (Institut de Chimie, Strasbourg). I am grateful to the facility users for their permission to show results obtained with their molecules. I thank A. Marquette and L. Vermeer (Institut de Chimie, Strasbourg) for having shared their knowledge about light scattering, and to I. Billas and P. Poussin-Courmontagne (IGBMC, Illkirch) for help with the SEC-MALS analysis.

## References

[1] P. Debye Molecular-weight determination by light scattering, J. Phys. Colloid Chem. (1947) 51 (1) 18-22.

[2] B.H. Zimm, The scattering of light and the radial distribution function of high polymer solutions, J. Chem. Phys (1948) 16 (12) 1093-1099.

[3] B. Chu, Laser light scattering, Academic Press, New York, 1974.

[4] R. Pecora, Dynamic light scattering. Applications of photon correlation spectroscopy, Plenum Press New York, 1985.

[5] K.S. Schmitz, An introduction to Dynamic light scattering by macromolecules, Academic Press New York, 1990.

[6] D. Oberthuer, E. Melero-Garcia, K. Dierks, A. Meyer, C. Betzel, A. Garcia-Caballero, J.A. Gavira J., Monitoring and scoring counter-diffusion protein crystallization experiments in capillaries by in situ dynamic light scattering. PlosOne 7 (2012) e33545.

[7] J.A. Finnegan, D.J. Jacobs, Light scattering from benzene, toluene, carbon disulphide and carbon tetrachloride. Chem. Phys. Lett. (1970) 6 (3) 141-143.

[8] H.D.T. Mertens, D.I. Svergun, J. Struct. Biol. (2010) 172, 128-141. Structural characterization of proteins and complexes using small-angle X-ray solution scattering.

[9] D.E. Koppel, Analysis of macromolecular polydispersity in intensity correlation spectroscopy: The method of cumulants. The Journal of Chemical Physics (1972) 57 (11) 4814-4820.

[10] S.W. Provencher, Contin: a general purpose constrained regulatization program for inverting noisy linear algebraic and integral equations. Computer Physics Comm. (1982) 27, 229-242.

[11] R. C. Weast (Ed.) Handbook of Chemistry and Physics CRC Press, Cleveland OH, 55[th] ed., 1974-75.

[12] P. Claes, M. Dunford, A. Kenney, P. Vardy, An online dynamic light-scattering instrument for macromolecular characterisation, in: S. Harding, D. Satelle and V. Bloomfield (Eds.), Laser scattering in Biochemistry, London: Royal Society of Chemistry, 1992, 66-76.

[13] A. Gobert, F. Pinker, O. Fauhsbauer, B. Gutmann, R. Boutin, P. Roblin, C. Sauter, P. Giegé, Structural insights into protein-only RNase P complexed with tRNA, Nat. Commun. (2013) 4, 1353.

[14] D. Destouches, N. Page, Y. Hamma-Kourbali, V. Machi, O. Chaloin, S. Frechault, C. Birmpas, P. Katsoris, J. Beyrath, P. Albanese, M. Maurer, G. Carpentier, J.M. Strub, A. Van Dorsselaer, S. Muller, D. Bagnard, J. P. Briand, J. Courty, A simple approach to cancer therapy afforded by multivalent pseudopeptides that target cell-surface nucleoproteins, Cancer Res. (2011) 71, 3296 -3305.

[15] N. Page, N. Schall, J.M. Strub, M. Quinternet, O. Chaloin, M. Décossas, M.T. Cung, A. Van Dorsselaer, J.P. Briand, S. Muller, The spliceosomal phosphopeptide P140 controls the lupus disease by interacting with the HSC70 protein and via a mechanism mediated by gamma delta T cells. PLoS One. (2009) 4 (4) e5273.

[16] C. Zinglé, D. Tritsch, C. Grosdemange-Billiard, M. Rohmer, Cathecol-rhodanine derivatives: Specific and promiscuous inhibitors of Escherichia coli deoxyxylulose phosphate reductoisomerase (DXR), Bioorganic & Medicinal Chemistry (2014à 22, 3713-3719.

[17] B. Lorber, J.B. Bishop, LJ. DeLucas, Purification of octyl-β-D-glucopyranoside and re-estimation of its micellar size. Biochem. Biophys. Acta (1990) 1023, 254-265.

[18] P. Schellenberger, G. Demangeat, O. Lemaire, C. Ritzenthaler, M. Bergdoll, V. Oliéric, C. Sauter, B., Strategies for the crystallization of viruses: Using phase diagrams and gels to produce 3D crystals of Grapevine Fan Leaf Virus. J. Structural Biol. (2011) 174, 344-351.

[19] Lorber B. (2017) Better results with homogeneous biological macromolecules. HAL-01649384, https://hal.archives-ouvertes.fr/hal-01649384.

[20] L. Bonnefond, M. Frugier, E. Touzé, B. Lorber, C. Florentz, R. Giegé, J. Rudinger-Thirion, C. Sauter C., Tyrosyl-tRNA synthetase: the first crystallization of a human mitochondrial aminoacyl-tRNA synthetase, Acta Cryst. F (2007) F63, 338-341.

[21] A. Neuenfeldt, B. Lorber, E. Ennifar, A. Gaudry, C. Sauter, M. Sissler, C. Florentz, Thermodynamic properties distinguish human mitochondrial aspartyl-tRNA synthetase from bacterial homolog with same 3D architecture, Nucleic Acids Res. (2013) 41, 2698-2708.

[22] B. Lorber, M. Adrian, J. Witz, M. Erhardt, R.J., Formation of two-dimensional crystals of icosahedral RNA viruses, Micron (2008) 39, 431-446.

[23] M. Carrillo-Tripp, C. M. Shepherd, I. A. Borelli, S. Venkataraman, G. Lander, P. Natarajan, J. E. Johnson, C. L. Brooks, III , V. S. Reddy, VIPERdb2: an enhanced and web API enabled

relational database for structural virology, Nucleic Acid Res. 37 (2009) D436-D442; http:/*viperdab.scripps.edu/*

[24] T. Salinas, S. El Farouk-Ameqrane, E. Ubrig, C. Sauter, A.M. Duchêne, L. Maréchal-Drouard, Molecular basis for the differential interaction of plant mitochondrial VDAC proteins with tRNAs, Nucleic Acids Res. (2014) 42 (15) 9937-48.

[25] R.P. Gonçalves, N. Buzhynskyy, V. Prima, J.N. Sturgis, S. Scheuring,Supramolecular assembly of VDAC in native mitochondrial outer membranes, J. Mol. Biol. (2007) 369 (2) 413-418.

[26] J. Lipfert, L. Columbus, V.B. Chu, S.A. Lesley, S. Doniach, Size and shape of detergent micelles determined by small-angle x-ray scattering J. Phys. Chem. B (2007) 111, 12427-12438.

[27] P. Wolff, I. Amal, V. Olieric, O. Chaloin, G. Gygli, E. Ennifar, B. Lorber, G. Guichard, J. Wagner, A. Dejaegere, D. Burnouf, Differential modes of peptide binding onto replicative sliding clamps from various bacterial origins. J. Med. Chem. (2014) 57, 7565-7576.

[28] R. Li, L M. Macnamara, J. D. Leuchter, R. W. Alexander, S. S. Cho, MD Simulations of tRNA and Aminoacyl-tRNA Synthetases: Dynamics, Folding, Binding, and Allostery, Int. J. Mol. Sci. (2015) 16, 15872-15902.

[29] C. Hemmer, I. Orlov, L. Ackerer, A. Marmonier, K. Hleibieh, C. Schmitt-Keichinger, E. Vigne, S. S. Gersch, V. Komar, L. Belval, F. Berthold, B. Monsion, P. Bron, O. Lemaire, B. Lorber, C. Gutiérrez, S. Muyldermans, G. Demangeat, B.P. Klaholz, C. Ritzenthaler. Cryo-EM structure of a nanobody-bound plant virus reveals molecular basis of resistance and vector transmission, Proceedings of the 16th European Microscopy Congress (2016) Lyon. Abstract 5807, DOI: 10.1002/9783527808465.EMC2016.5807.

[30] M. Bailly, M. Blaise, H. Roy, M. Deniziak, B. Lorber, C. Birck, H.D. Becker, D. Kern, tRNA-dependent asparagine formation in prokaryotes. Characterization, isolation and structural and functional analysis of a ribonucleoprotein particle generating Asn-tRNA$^{Asn}$, Methods (2008) 44, 146-163

[31] M. Blaise, M. Bailly, M. Frechin, M. A. Behrens, F. Fischer, C. L. P. Oliveira, H. D. Becker, J. S. Pedersen, S. Thirup, D. Kern, Crystal structure of a transfer-ribonucleoprotein particle that promotes asparagine formation, EMBO J. (2010) 29, 3118–3129.

[32] F. Fischer, J. L. Huot, B. Lorber, G. Diss, T.L. Hendrickson, H. D. Becker, J. Lapointe, D. Kern, The asparagine-transamidosome from *Helicobacter pylori*: a dual-kinetic mode in non-discriminating aspartyl-tRNA synthetase safeguards the genetic code, Nucleic Acids Res. (2012) 40 (11) 4965–4976.

[33] S.N. Timasheff, M.J. Kronman, The extrapolation of light-scattering data to zero concentration, Arch. Biochem. Biophys. (1959) 83, 60-75.

[34] C. Sauter, B. Lorber, A. Gaudry, L. Kari, H. Schwenzer, F. Wien, P. Roblin, C. Florentz, M. Sissler, Neurodegenerative disease-associated mutants of a human mitochondiral aminoacyl-tRNA synthetase present individual molecular signatures, Nature Sci. Rep. (2015) 5, 17332.

[35] B. Raynal, P. Lenormand, B. Baron, S. Hoos, P. England, Quality assessment and optimization of purified protein samples: why and how? Microbial Cell Factories (2014) 13, 180. DOI 10.1186/s12934-014-0180-6.

[36] R. Kurapati, J. Russier, M.A. Squillaci, E., C. Ménard-Moyon, A.E. Del Rio-Castillo, E. Vazquez, P. Samorì, V. Palermo, A. Bianco, dispersibility-dependent biodegradation of graphene oxide by myeloperoxidase, Small (2015) 11 (32) 3985-94.

[37] A. Marquette A., B. Lorber, B. Bechinger, Reversible liposome association induced by LAH4: a peptide with potent antimicrobial and nucleic acid transfection activities. Biophys. J. (2010) 98, 2544-2553.

[38] D.J. Gibbings, C. Ciaudo, M. Ehrhardt, O.Voinnet, Multivesicular bodies associate with components of miRNA effector complexes and modulate miRNA activity, Nat. Cell. Biol., (2009) 11 (9), 1143-1149.

[39] A.D. Hanlon, M.I. Larkin, R.M. Reddick, Free-solution, label-free protein-protein interactions characterized by dynamic light scattering, Biophys. J. (2010) 98, 297-304.

[40] D. Roberts, R. Keeling, M. Tracka, C.F. van der Walle, S. Uddin, J. Warwicker, R. Curtis, The role of electrostatics in protein-protein interactions of a monoclonal antibody, Mol. Pharm. (2014) 11, 2475-2489.

[41] E.N. Lewis, W. Qi, L.H. Kidder, S. Amin, S.M. Kenyon, S. Blake, Combined dynamic light scattering and Raman spectroscopy approach for characterizing the aggregation of therapeutic proteins, Molecules (2014) 19 (12), 20888-20905.

[42] A.P. Minton, Recent applications of light scattering measurement in the biological and pharmaceutical sciences, Anal. Biochem. (2016) 501, 4-22.

**Supplementary material**

*Corrections for sample properties in SLS*

The computation of particle mass from Eq. 3 requires the increment of refractive index (*dn/dc*). Proteins with an average amino acid composition have a *dn/dc* in the range 0.18 - 0.19 mL/g. It may be necessary to determine the *dn/dc* of nucleic acids or of assemblies made of proteins and nucleic acids. An Abbe refractometer is useful to measure the refractive index on ten microliters solution at one or a few precisely known concentrations (e.g. 10 to 20 mg/mL calculated from measured absorbance and from extinction coefficient, either calculated from the chemical composition or determined experimentally).

*Corrections for solvent properties in DLS*

The refractive index *n* and the viscosity *η* of the solvent in Equations 9 to 11, vary with solvent composition and temperature. Appropriate corrections take into account these variations. The refractive index *n* expresses how much the solvent refracts a light beam. It is equal to the ratio of the speed of light in the medium over the speed of light in vacuum and varies with the wavelength of the light. *n* can be measured with an Abbe refractometer. For pure water, it is 1.333 at 20°C as compared to ~1.0003 for air at λ = 550 nm, and varies by only ~0.00009 per °C between 15°C and 25°C [11]. The viscosity reflects the resistance encountered by the diffusing particles. The kinematic viscosity (unit Stokes, 1 St = 1 cm$^2$.s$^{-1}$) is measured at constant temperature with an Anton-Paar AMV falling-bead viscosimeter. Multiplication by the solvent density converts it to dynamic (or absolute) viscosity *η* needed here. The units of viscosity are the centipoise, cP, and the milliPascal per second, so that 1 cP = 1 mPa·s). The density of the solvent (in g/L) is determined in three steps with a pyknometer, i.e. a little flask of ~5 or ~10 mL equipped with a stopper topped with a beveled glass tube. Assuming a density of 1 g/L for pure water at 20°C, the mass of the flask filled with pure water minus the weight of the empty flask gives its volume. The weight of the flask filled with solution divided by the volume yields the searched density. Viscosity varies significantly with temperature. For water, *η* = 1.002 cP at 20°C. It is 13.6% less at 15°C and 12.5% more at 25°C. Some chemicals increase viscosity while others diminish it. Glycerol (*M$_r$* ~92, d ~1.26 g/L, molarity 13.6 M) is an ingredient of many buffer solutions used with proteins. At 20°C, a 10% (m/v) or 1.36 M aqueous glycerol solution is ~1.37 times more viscous than water. Corrections for viscosity are meaningful only if the viscosity of the solvent is less than two, as in the case of the majority of buffer solutions containing less than 15 % (v/v) glycerol.

*Subtraction of solvent particle size distribution in DLS*

The conversion of the particle size distribution (PSD) by intensity to a PSD as a function of the real intensities (see the example in **Fig. 5**) is essential before subtracting the particle size distribution (PSD) of the solvent from that of the macromolecule. Both operations are mandatory when the solvent contains large particles such as detergent micelles.

*Importance of temperature*

A pH variation may accompany a temperature variation depending on the $\Delta pk_a/°C$ of the buffering molecule. This variation may then alter the charge, the conformation and the solubility of the biological macromolecule under study. For this reason, it is always better to compare results obtained the same temperature with various methods.

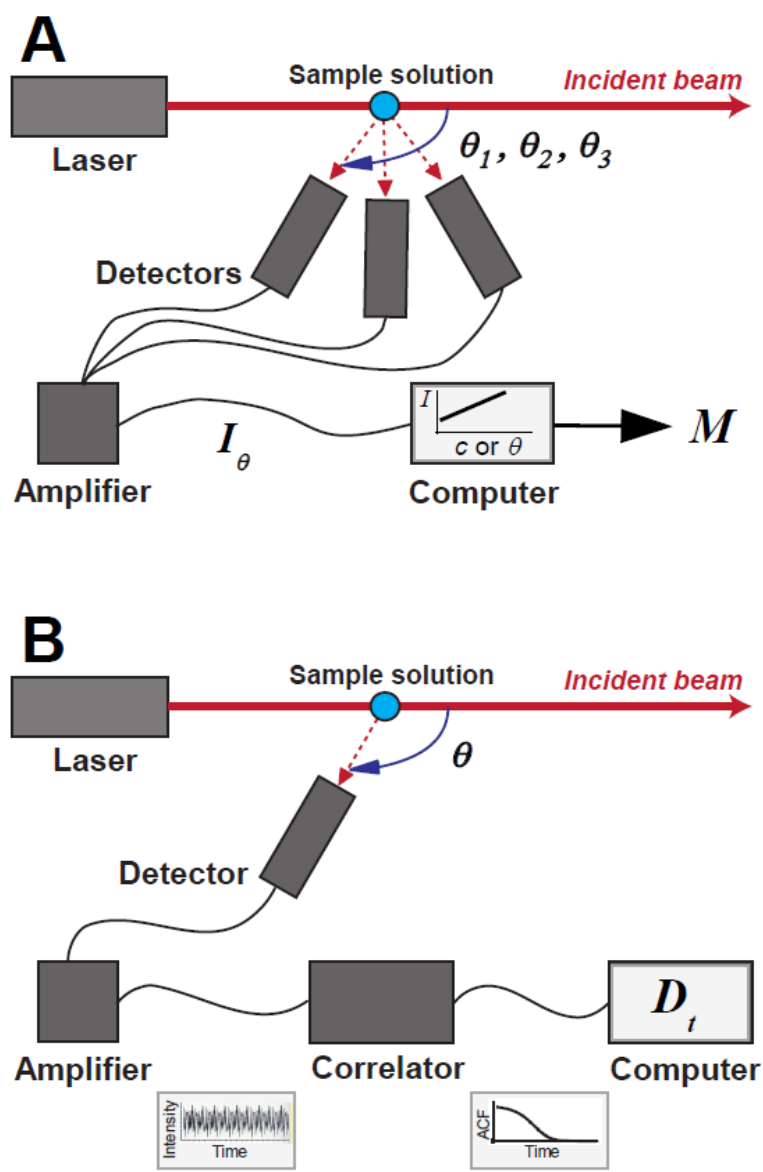

**Fig. 1: Components of SLS and DLS systems.** (A) SLS records the time-averaged intensity of the scattered light to extract the absolute particle mass. (B) DLS uses the time-dependent fluctuations of the scattered light due to Brownian motion to derive a diffusion coefficient and calculate the radius of the hard sphere that has the same diffusion coefficient. The angle $\theta$ is measured clockwise starting from the incident beam that has traversed the sample.

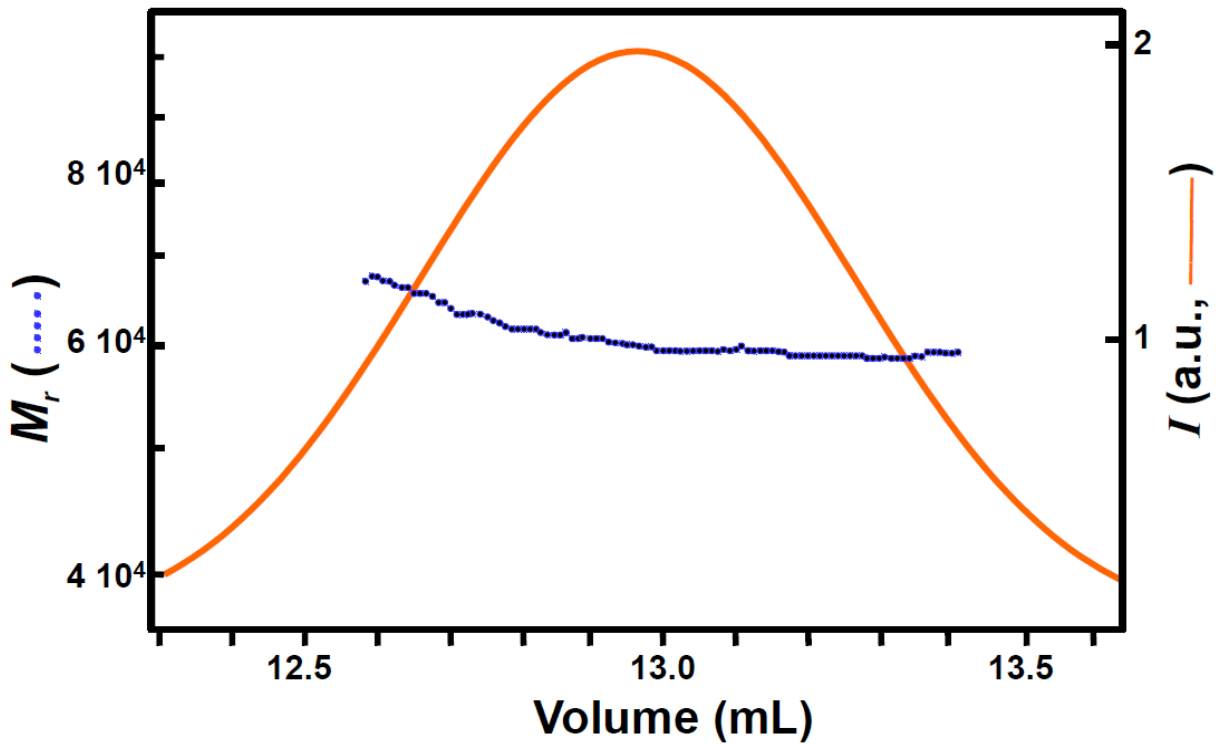

**Fig. 2: SEC-MALS analysis of a protein.** Two hundred microgram of *Arabidopsis thaliana* Protein Only RNase P-2 (PRORP-2) dissolved in 100 microliters buffer solution (containing 50 mM Hepes-Na, 250 mM NaCl, 5% (v/v) glycerol and 1.6 mM TCEP) are loaded onto the SEC column and eluted with the same mobile phase. The instrumentation records the scattered intensity together with UV absorbance at 280 nm and refractive index. The plot shows the variation of absorbance as a function of the elution volume. Across the absorbance peak, the calculated $M_r$ is ~60,000. The mass recovery is 28% in the absorbance peak and the rest of the protein distributes over large aggregates that elute at smaller volumes (not shown).

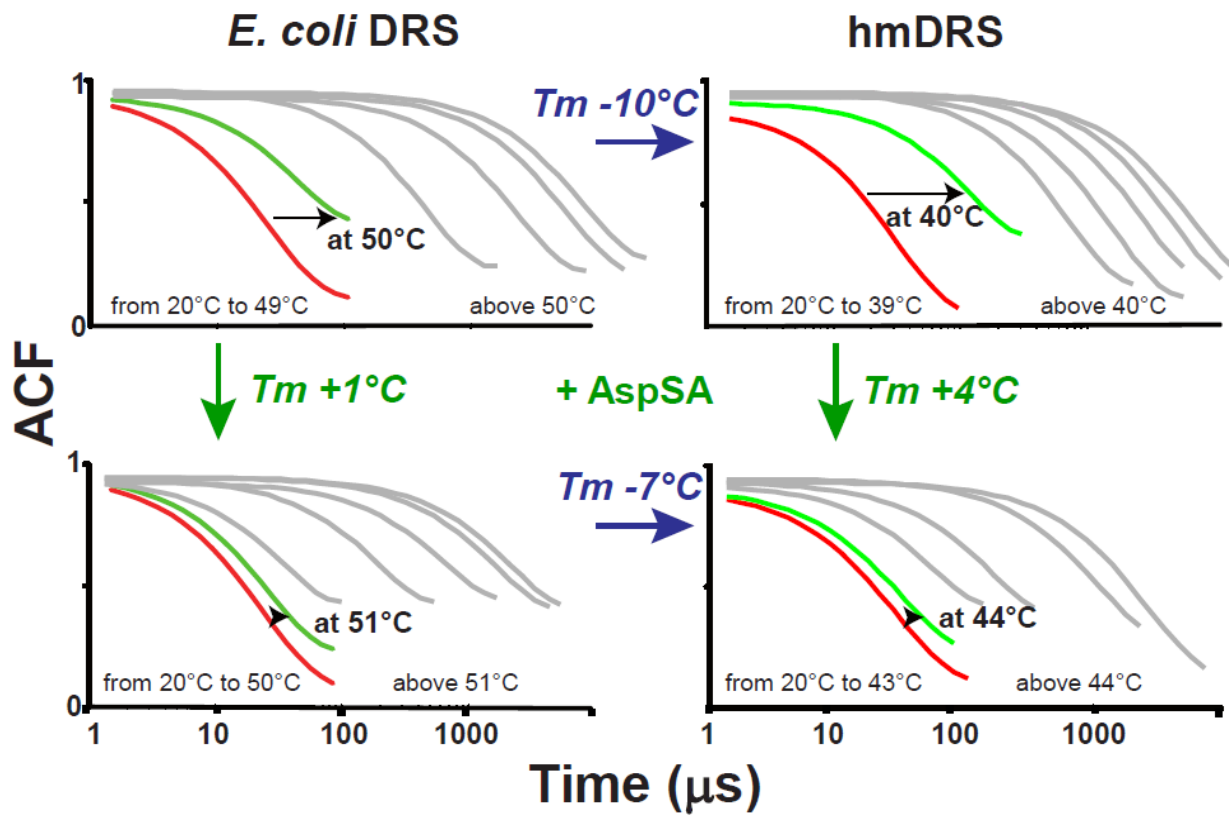

**Fig. 3: Protein thermal denaturation monitored by DLS.** *E. coli* and human mitochondrial (hm) aspartyl-tRNA synthetase (DRS) at 3 mg/mL in 50 mM HEPES-Na pH 7.5, 150 mM NaCl, 1 mM DTT, 0.1 mM EDTA and 10% (v/v) glycerol are heated from 20°C to 80°C in the Zetasizer. For the sake of clarity, the figure displays only a part of the distribution fits. The shift towards greater times indicates the formation of large size scatterers (i.e. protein aggregates) consecutive to protein subunit dissociation and unfolding. Melting occurs at higher temperatures ($T_m$) in the presence of the ligand aspartyl-sulfamoyl ATP (AspSA).

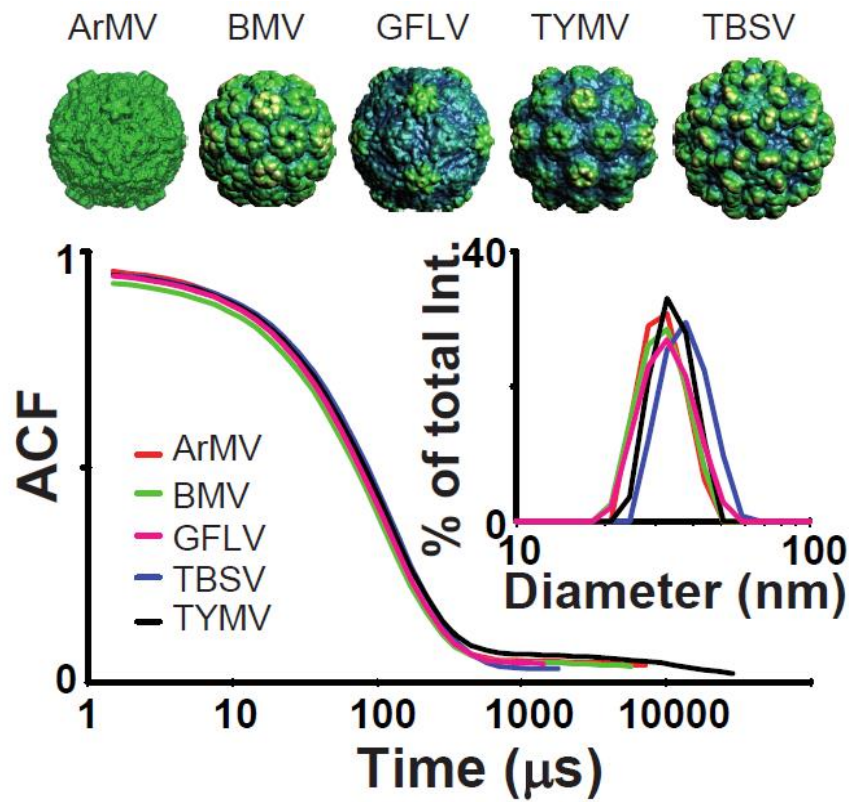

**Fig. 4: Size of icosahedral plant viruses.** ACFs and PSDs (in the inset) of five viruses at 0.05 mg/mL in 150 mM NaCl solution analyzed separately at 20°C in the Zetasizer. The images of the viral capsids are prepared using the VIPERdb software [23] with Protein Data Bank files for Brome Mosaic Virus (BMV, 1js9), Grapevine Fan Leaf Virus (GFLV, 4v5t), Tomato Bushy Stunt Virus (TBSV, 2tbv), and Turnip Yellow Mosaic Virus (TYMV, 1auy). The image of the *Arabis* Mosaic Virus (ArMV) capsid is the cryo-electron microscopy envelope (Electron Microscopy Data Bank at the European Bioinformatics Institute, accession code EMD-2242). Mean hydrodynamic diameters: ArMV, ~32 nm; BMV, ~32 nm; GFLV, ~33 nm; TBSV, ~37 nm and TYMV, ~34 nm.

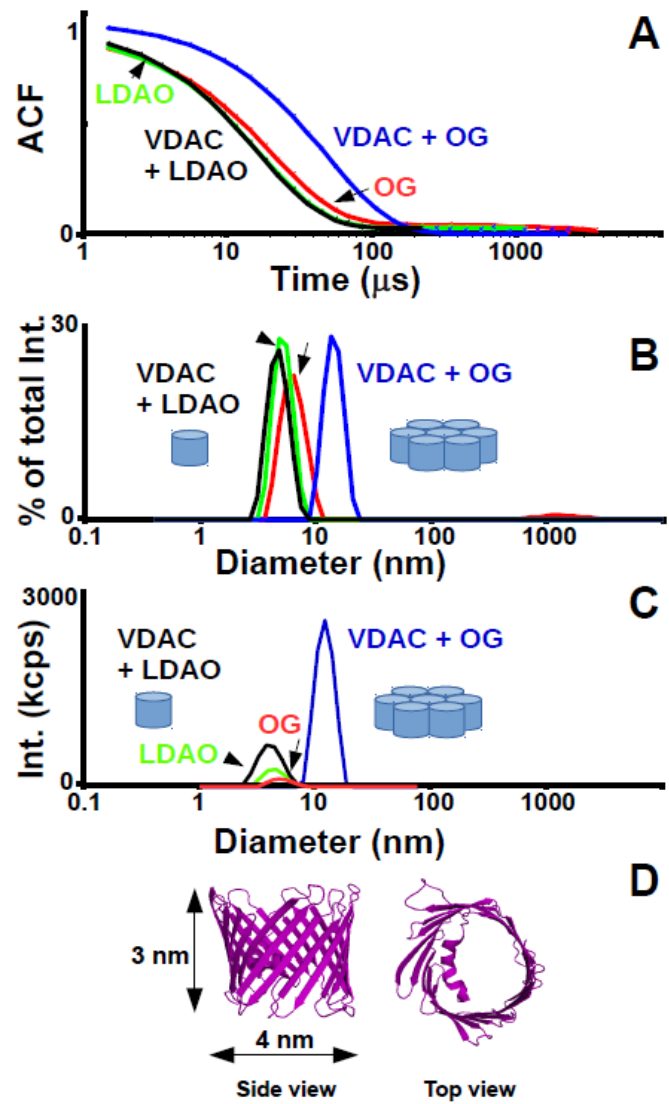

**Fig. 5: Oligomeric structure of a membrane protein.** (A) ACFs of VDAC 34 solubilized in octyl glucoside (OG) and lauryldimethylamine oxide (LDAO) and of each detergents alone. The protein is at 1.4 mg/mL in 10 mM sodium phosphate and 50 mM sodium sulfate pH 7.5 (for the solvent without detergent, $n$ = 1.333 and $\eta$ = 1.119 cP at 20°C). The DLS analyses of detergents are at 1% (m/v) performed in the Zetasizer at 20°C. (B) PSDs as a function of intensity assuming the total intensity is equal to 100%. (C) PSDs taking into account the mean total intensity of the sample (9400 kcps for VDAC in the presence of OG and 2250 kcps for VDAC in the presence of LDAO) but not corrected for the solvent. The diameters of the protein-detergent complexes are 4.7 ± 2 nm and 14.5 nm ± 2 nm in LDAO and in OG, respectively. The drawings represent the VDAC monomer and the hexagonal packing. (D) Crystallographic structure of human VDAC (PDB, emn).

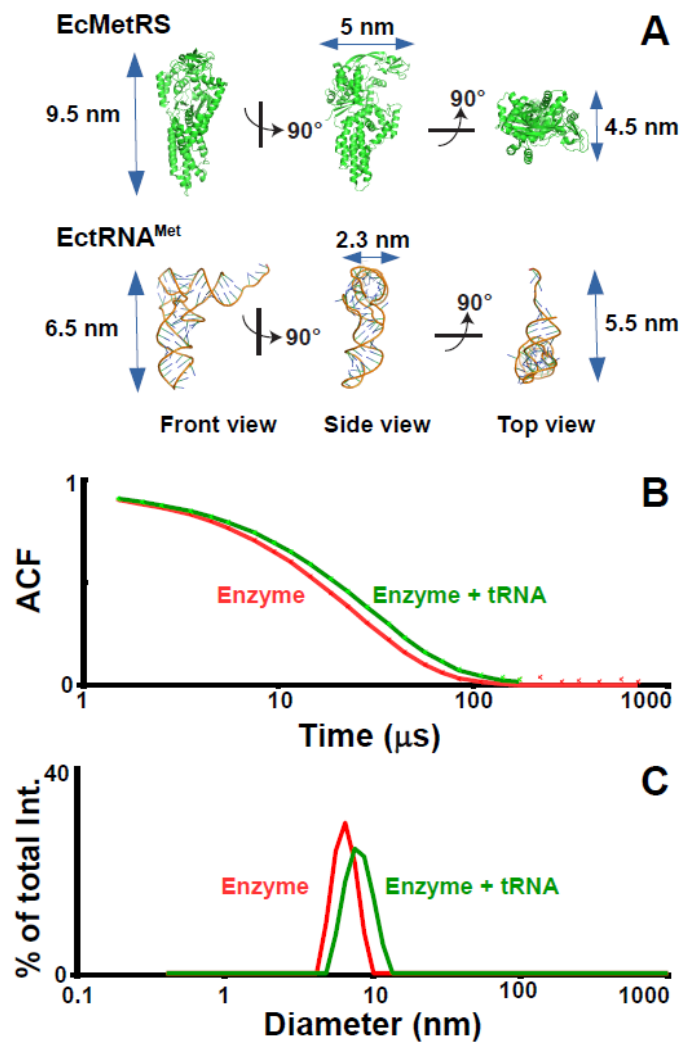

**Fig. 6: Association of a protein and a nucleic acid.** The monomeric domain of *E. coli* methionyl-tRNA synthetase (EcMetRS, $M_r$ 64,000) is analyzed at 20°C in the Zetasizer, in the absence and in the presence of a 10% excess of *E. coli* tRNA$^{Met}$ (EctRNA$^{Met}$, $M_r$ 25,000) molecules. The protein is at 5 mg/mL in 20 mM Tris pH 7.5 and 200 mM NaCl ($n$ = 1.332, $\eta$ = 1.014 cP) after 1 h ultracentrifugation at 100,000 x g. (A) 3D crystallographic structures of EcMetRS (PDB, 1qqt) and tRNA$^{Met}$ from yeast (PDB,1yfg). (B) ACFs and (C) PSDs of free enzyme and enzyme/tRNA complex. The mean $d_h$ increases from 6.5 ± 1 nm to 8 ± 1 nm after saturation with the ligand.

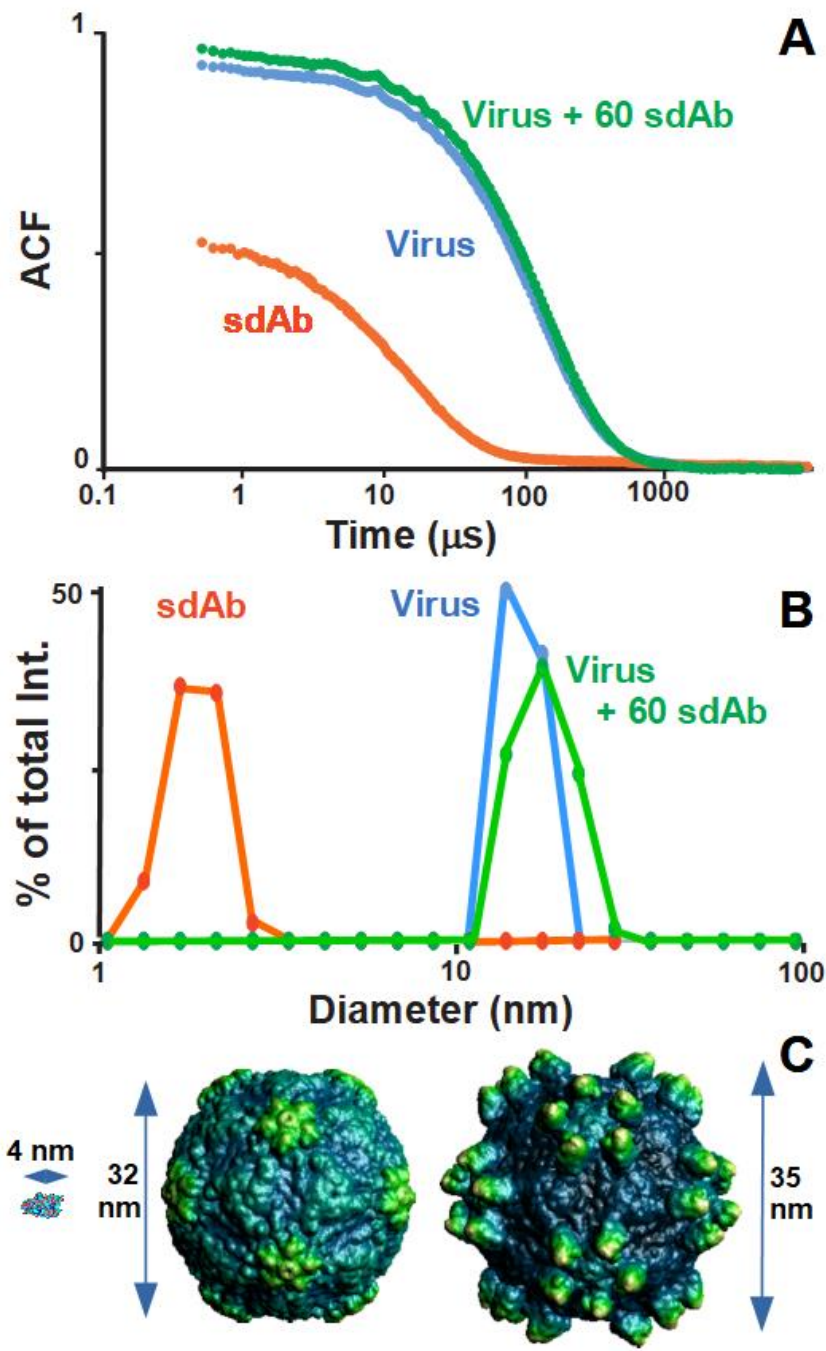

**Fig. 7: Titration of a virus with a protein:** Grapevine Fan Leaf Virus is titrated with a single domain antibody (sdAb) to a stoichiometry of 100, i.e. beyond full saturation of the sixty copies of the capsid protein. (A) ACFs and (B) PSDs of sdAb, virus and coated virus. The DLS measurements are done at 20°C in the Nanostar on sdAB at 1 mg/mL, virus at 0.1 mg/mL in 150 mM NaCl ($n = 1.332$, $\eta = 1.018$). (C) 3D crystallographic structures of the antibody (PDB, 5foj), the virus (PDB, 4v5t) and the complex (PDB, 5foj) at scale (from left to right).

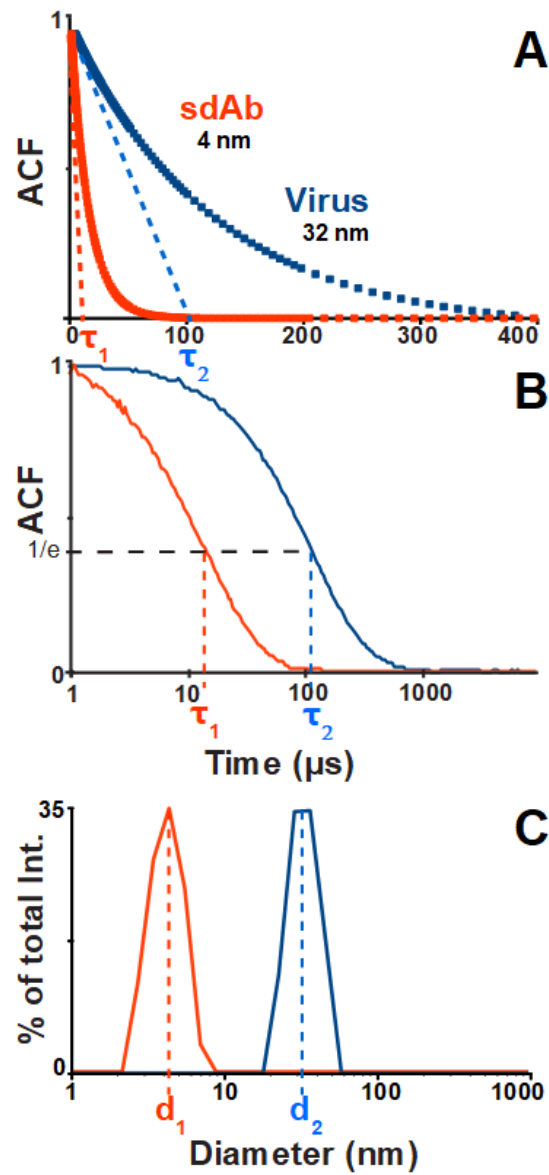

**Supplementary Fig. 1: Autocorrelation functions and particle size distribution.** (A,B) Autocorrelation functions (ACFs) and (C) article size distributions (PSDs) of pure single domain camel antibody (sdAb, $M_r$ ~15000, c = 1 mg/mL in water) and of icosahedral Grapevine Fan Leaf Virus (c = 0.05 mg/mL in water) analyzed separately at 20°C in the Nanostar. The hydrodynamic diameters $d_h$ of the particles are ~4 nm and ~33 nm, respectively. The exponential ACFs show that both populations of particles are homogeneous. Panels A and B show how the delay time ($\tau_1$ ~10 μs for the small antibody and $\tau_2$ ~100 μs for the virus) is derived from ACF. Using Eq. 10, $D$ ~9 10$^{-6}$ cm$^2$/s and $D$ ~1.1 10$^{-6}$ cm$^2$/s, respectively. In (A) and (B), the time axes are in μs. Panel B displays only the useful part of the log$_{10}$ scale from 10$^{-10}$ s to 0.1 μs. (C) PSDs as a function of total intensity.

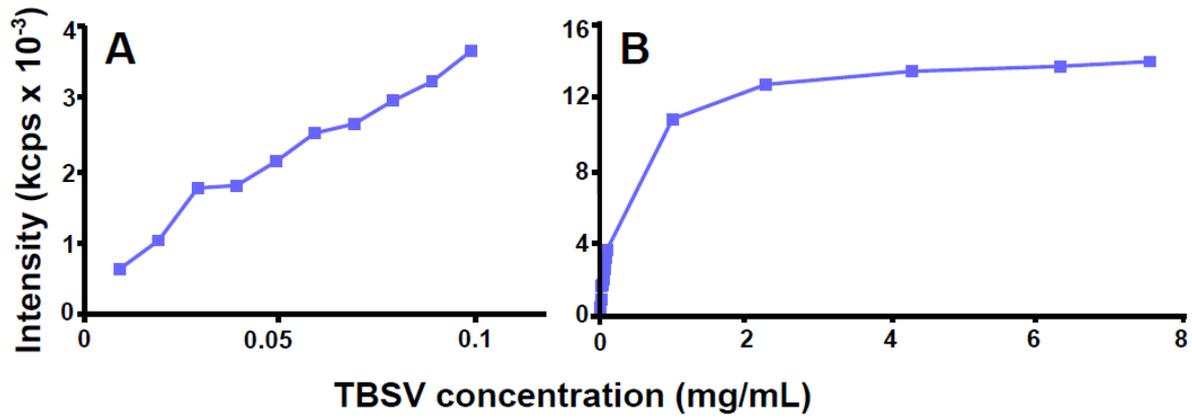

**Supplementary Fig 2: Detector response of a DLS instrument.** Variation of scattered intensity as a function of the concentration of Tomato Bushy Stunt Virus in water at 20°C. (A) Low and (B) high concentration range. The response of the detector is proportional to virus concentration only between 0.01 mg/mL to 0.1 mg/mL.

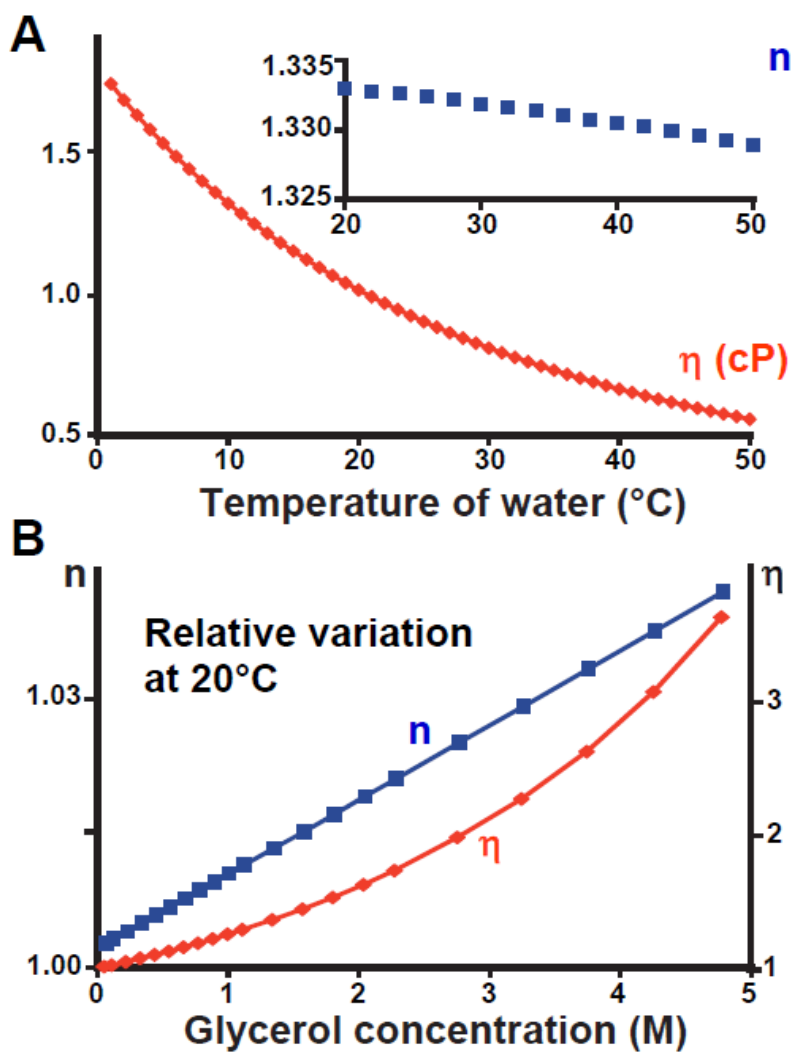

**Supplementary Fig 3**: **Viscosity and refractive index of water and of glycerol.** (A) Variation of the refractive index (without unit) and the viscosity of water (in cP) as a function of temperature. As a reference, for pure water $n = 1.333$ and $h = 1.002$ cP at 20°C. Chemicals dissolved in water shift the plots toward either lower or higher values. (B) Variation of the refractive index and of the relative viscosity of glycerol with molarity at 20°C. An aqueous 10% (v/v) glycerol solution has a molarity of 1.36 M. For data on others aqueous salt solutions, see [11].

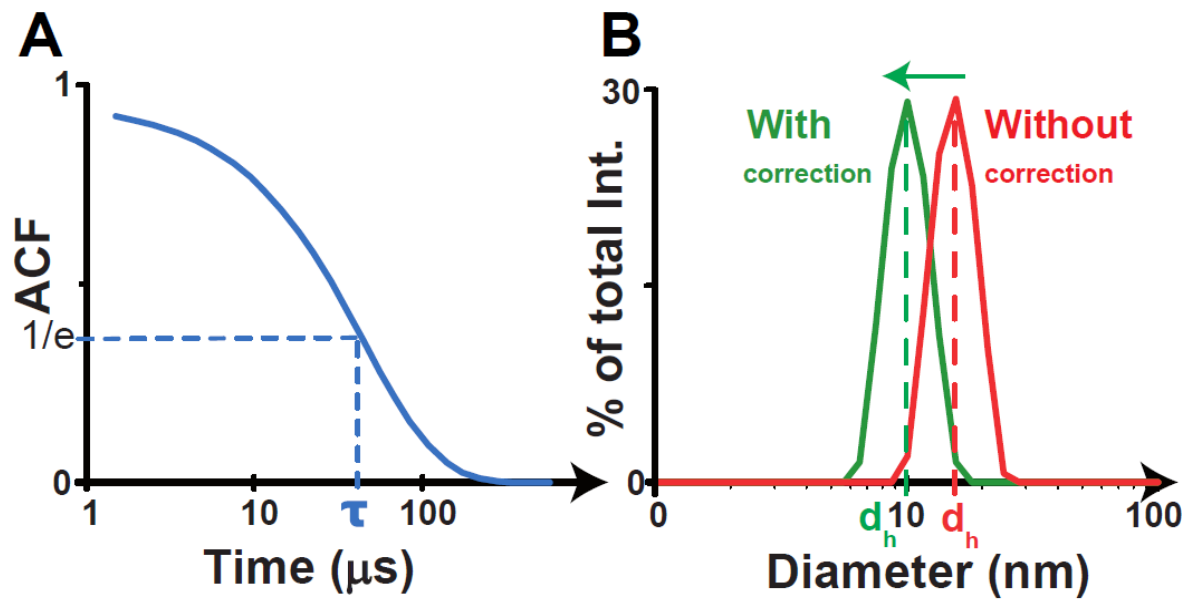

**Supplementary Fig 4: Effects of solvent properties on apparent protein size.** *E. coli* aspartyl-tRNA synthetase (20 μL at 1.2 mg/ml in 50 mM Hepes pH 7.5, 150 mM NaCl, 10% v/v glycerol, 1 mM DTT and 0.1mM EDTA) analyzed in the Zetasizer. The solvent has a refractive index $n = 1.352$ and an absolute viscosity $\eta = 1.49$ cP at 20°C, as compared to $n = 1.333$ and $\eta = 1.002$ cP for water. (A) ACF. (B) PSDs before and after corrections for solvent $n$ and $\eta$. The dissolved particles appear to be larger in the absence of correction because they diffuse more slowly in the viscous solvent. The correction decreases the $d_h$ from ~14 nm to ~10 nm and the $M_r$ of the equivalent globular protein from 320,000 to 145,000. This demonstrates that uncorrected data may result in wrong oligomeric structures.

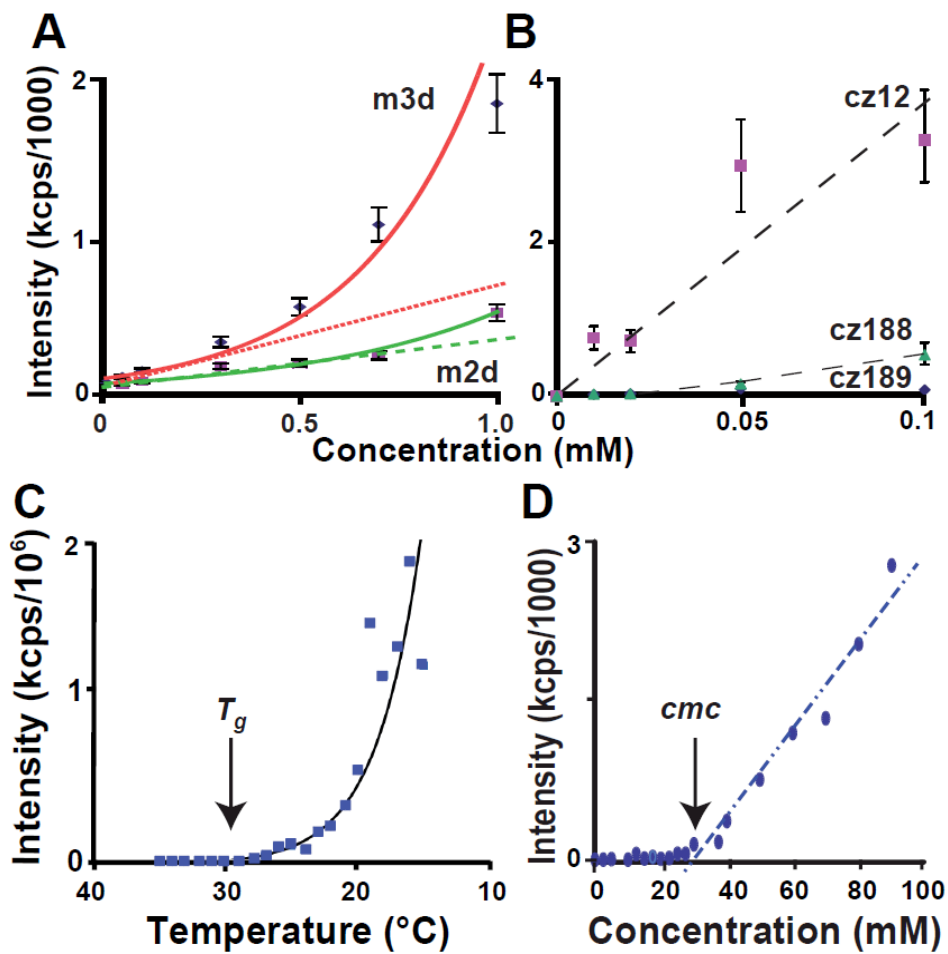

**Supplementary Fig 5: Aggregation phenomena.** (A) Plots of scattered *I* as a function of the concentration of peptides m2d and m3d in water. *I* values are means of five measurements with 10% error bar. Departure from linearity indicates the beginning of insolubility. (B) Variation of the intensity of scattered light as a function of the concentration of catechol-rhodanine derivatives solubilized in dimethyl sulfoxide and diluted in 50 mM Tris-HCl pH 7.5. The compounds cz188 and cz189 are much more soluble than compound cz12 that is insoluble already at c = 10 μM. (C) Variation of *I* with *T* during the gelling of a 0.4% (m/v) aqueous solution of agarose. (C) Scattered *I* as a function of octyl glucoside molarity. All data obtained with the Nanostar.

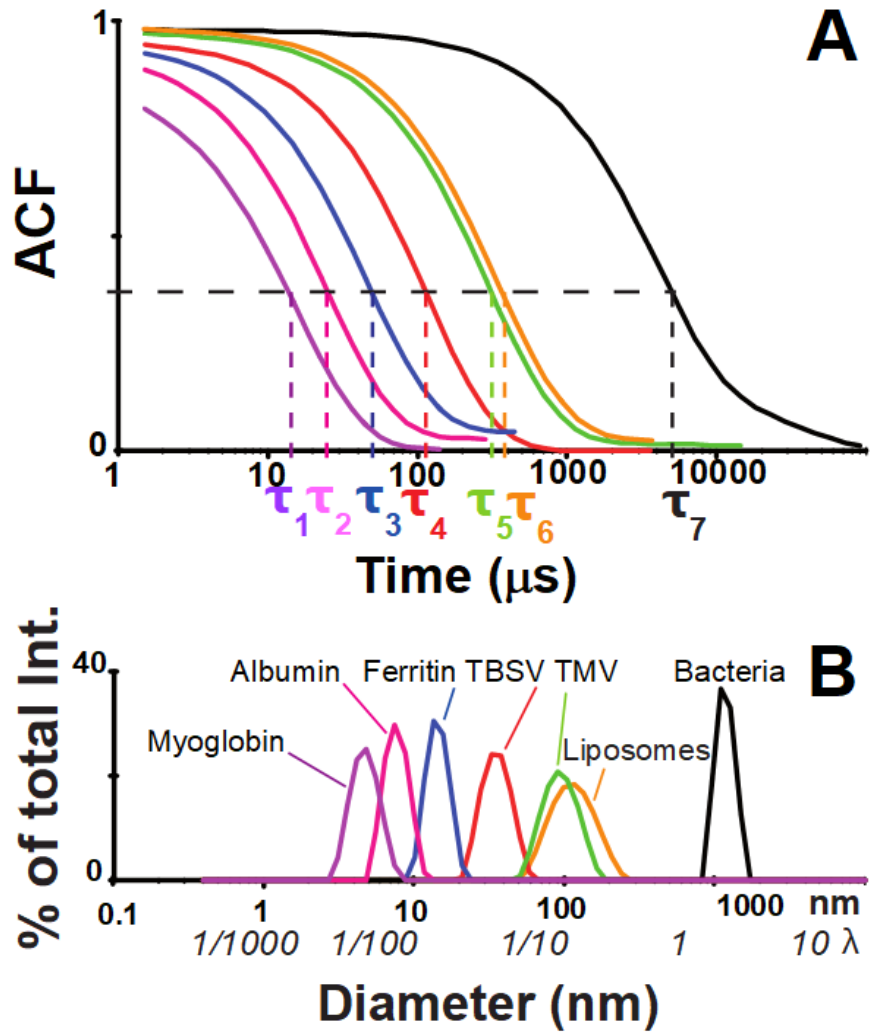

**Supplementary Fig. 6: Particle size range analyzable by DLS.** Overlay of (A) the ACFs and (B) the PSDs of three proteins, two viruses, liposomes and bacteria analyzed separately in the Zetasizer. (A) $\tau$ is the delay time of every particle population. In (B), the lower scale on the y-axis represents the fraction or multiple of the laser light wavelength ($\lambda$ = 633 nm). Samples are: sperm whale myoglobin ($M_r$ ~17,000, mean $d_h$ ~ 4,5 nm), bovine serum albumin ($M_r$ ~67,000, mean $d_h$ ~ 7,5 nm), horse spleen iron carrier ferritin ($M_r$ ~450,000, 24 subunit shell containing up to 4500 $Fe^{3+}$ ions, mean $d_h$ ~14 nm), icosahedral Tomato Bushy Stunt Virus ($M_r$ ~9 $10^6$, mean $d_h$ ~37 nm), rod-shaped Tobacco Mosaic Virus (length 150 to 300 nm, width ~18 nm, mean $d_h$ ~ 100 nm), liposomes (mean $d_h$ ~ 95 nm), and *Escherichia coli* cells (length up to 2 μm, width ~ 0.5 μm).

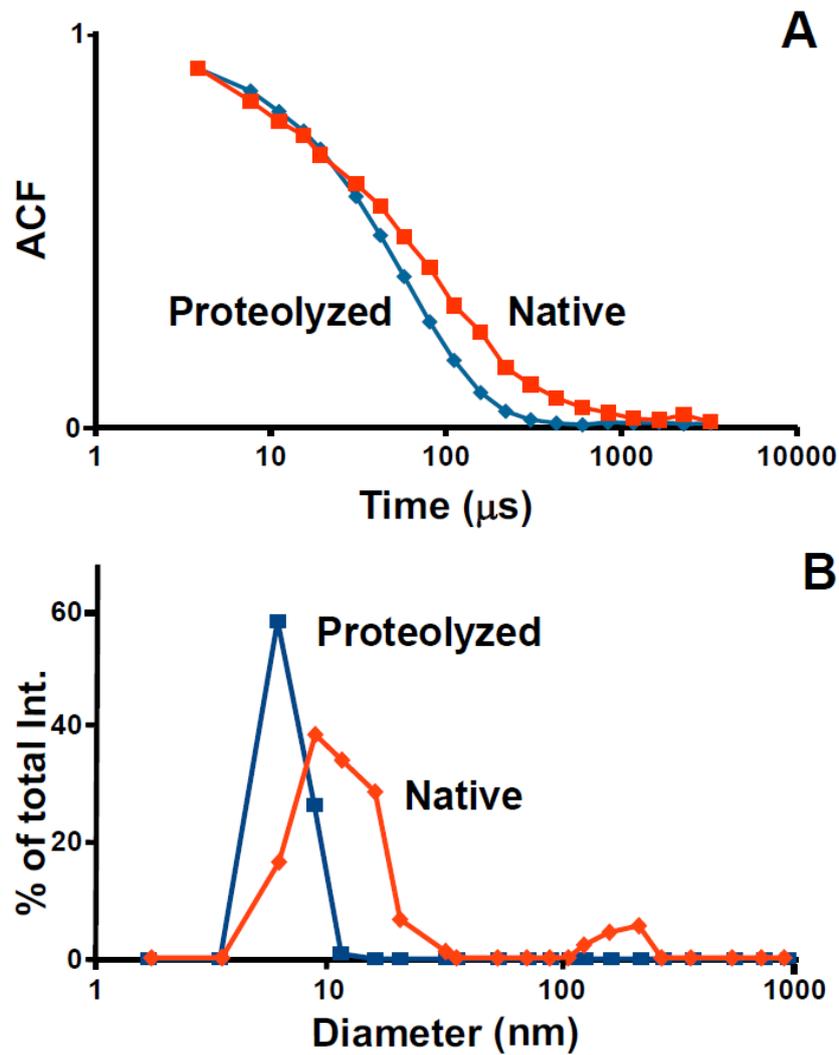

**Supplementary Fig. 7: Effect of limited proteolysis on protein homogeneity.** Human mitochondrial tyrosyl-tRNA synthetase analyzed by DLS in the DynaPro before and after limited trypsinolysis. (A) ACFs, (B) PSDs. The full-length enzyme (2 x 458 amino acid residues, $M_r$ ~103,000) is heterogeneous and polydisperse in 50 mM HEPES–NaOH pH 6.7, 300 mM NaCl and 10 mM DTE. It has a tendency to aggregate during handling and loses its activity above a concentration of 2 mg/mL. The major form has a mean $d_h$ ~11 ± 3 nm). At variance, the truncated protein is homogeneous ($d_h$ ~7 ± 1 nm).

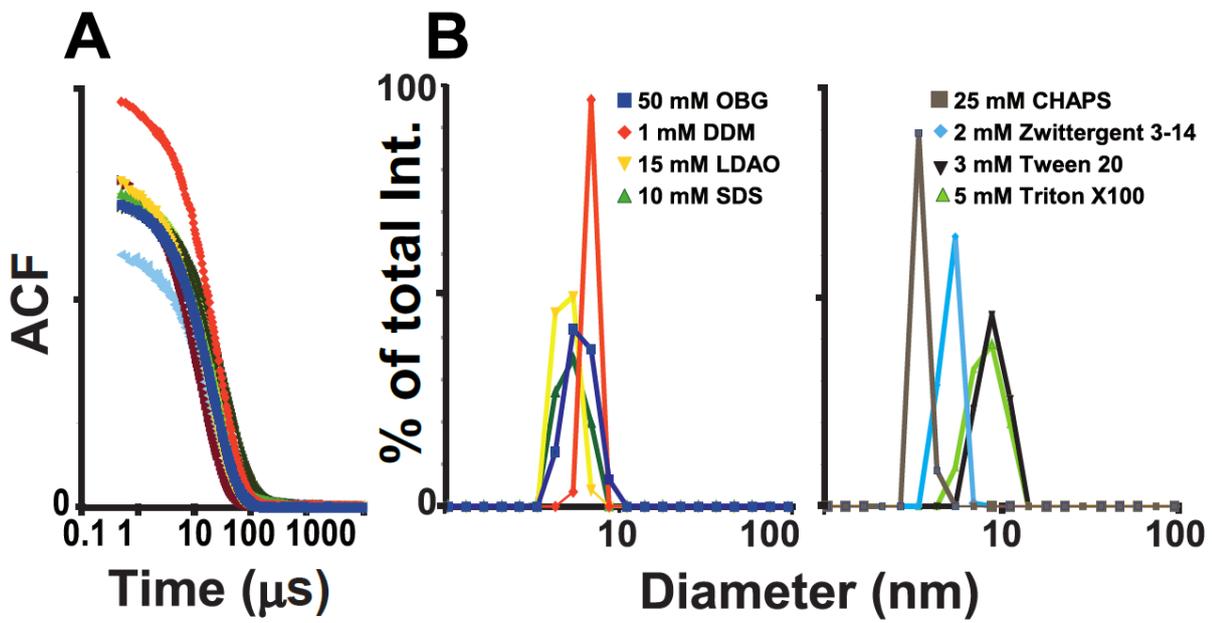

**Supplementary Fig. 8: Size of detergent micelles.** (A) ACFs and (B) PSDs of eight detergents dissolved in 150 mM NaCl and at concentrations above their cmc. Nanostar measurements performed at 20°C.

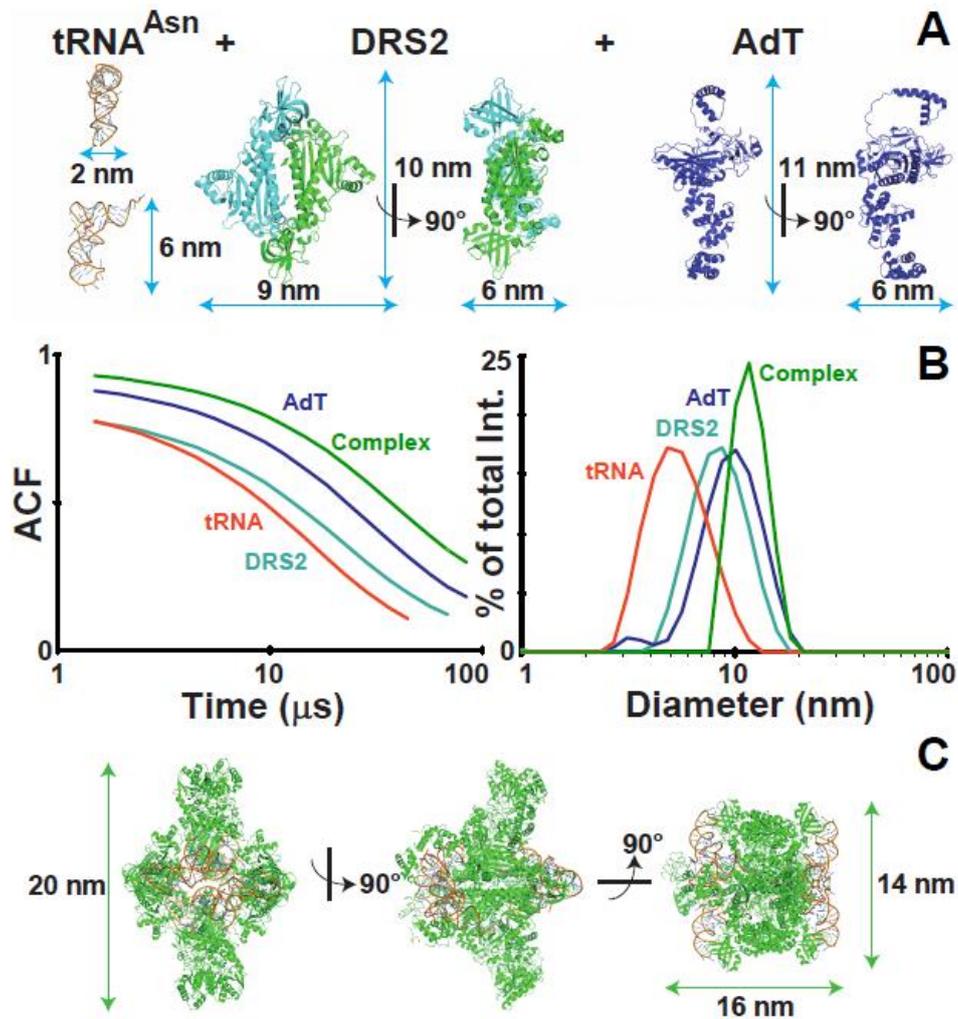

**Supplementary Fig. 9: Formation of a nucleoprotein complex.** (A) 3D crystallographic structures of tRNA[Asn], non-disciminating aspartyl-tRNA synthetase (DRS2) and amidotransferase (AdT) from *Thermus thermophilus.* (B) ACFs (left) and PSDs (right) of the three molecules and the ternary complex analyzed in the Zetasizer. (C) 3D Structure of the crystalline complex (PDB: 3kfu) containing an additional DRS2 dimer saturated by two tRNA[Asn] molecules. Notice that all molecules and the complex have shapes that are not spherical.

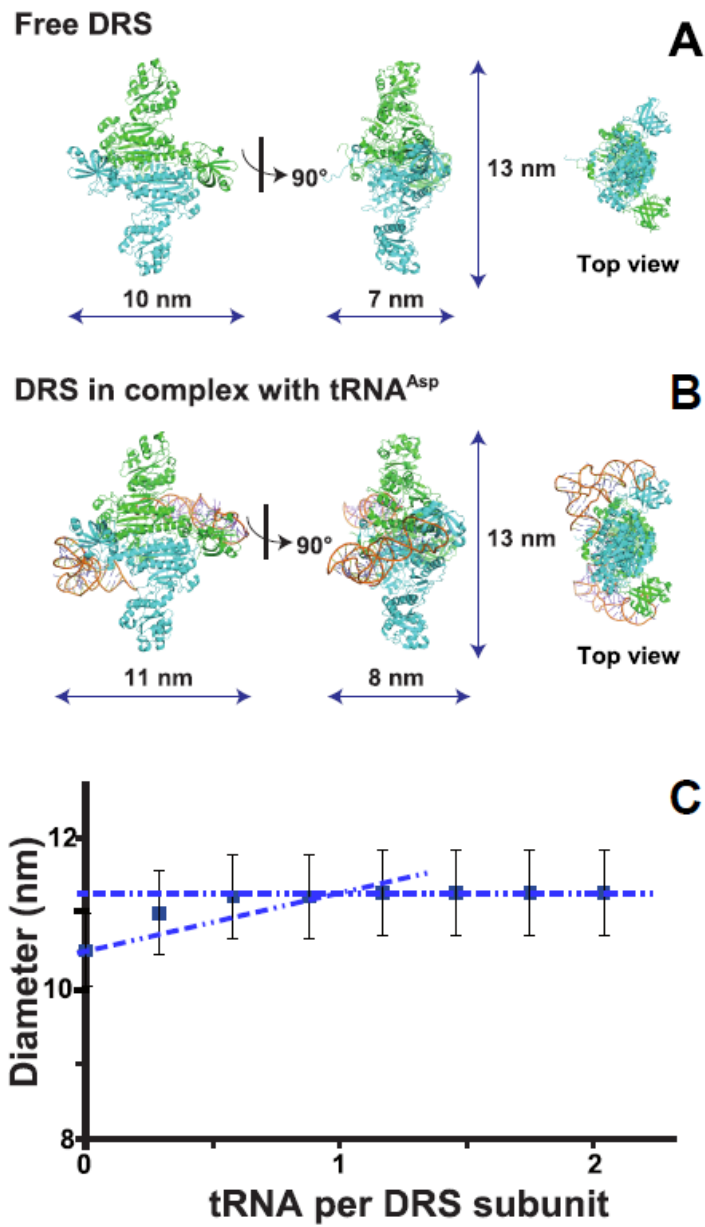

**Supplementary Fig. 10: Association of an ARN and a protein.** (A) 3D crystallographic structures and dimensions of free dimeric *E. coli* aspartyl-tRNA synthetase (DRS) and (B) of the complex with one tRNA$^{Asp}$ bound per protein subunit. Due to the complementarity of shapes, the length of the protein does not change but its other two dimensions increase by only one nm when it binds the tRNA. (C) Variation of the particle diameter with the ratio of tRNA$^{Asp}$ per DRS monomer of the homolog protein from human mitochondria. Analyses done in the Zetasizer with 0.5 mg/mL protein. Solvent is the same as in **Fig. 3**. This result is in marked contrast with that displayed in **Fig. 3**.

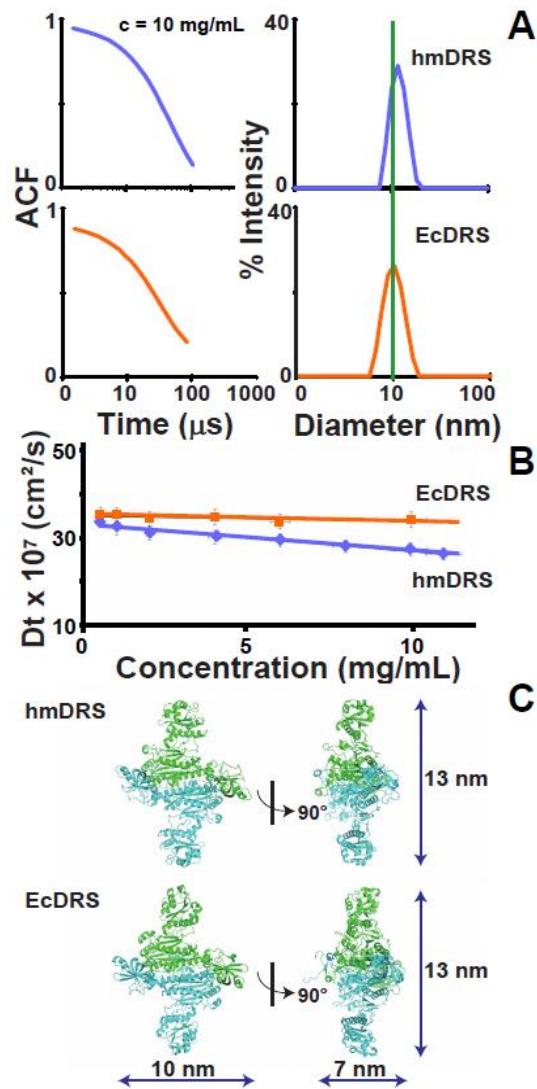

**Supplementary Fig 11: Interactions in protein solutions.** (A) ACFs (left) and PSDs (right) of aspartyl-tRNA synthetase from human mitochondria (hmDRS) deprived of its mitochondrial targeting sequence and of the homologous enzyme from *E. col*i (EcDRS). The human enzyme seems slightly greater than the bacterial one when analyzed at 25°C in the Zetasizer at 10 mg/mL in 50 mM HEPES-Na pH 7.5, 150 mM of NaCl, 1mM of DTT, 0.1 mM of ethylene diamine tetraacetic acid and 10% (v/v) glycerol ($n = 1.35$, $\eta = 1.35$ cP). (B) Variation of the diffusion coefficient $D$ (mean values with 5% error bars) of both proteins as a function of concentration. At high concentration, the $D$ of hmDRS is smaller, meaning that the enzyme behaves as a larger particle. Intermolecular interactions produce this effect since the mitochondrial enzyme actually has the same $D$ as the bacterial one at zero concentration. (C) The 3D crystallographic structures of hmDRS (PDB, 4ah6) and of EcDRS (PDB, 1eqr) confirm that both proteins have close dimensions.